\documentclass[twocolumn,times,tighten]{aastex631}

\usepackage{graphicx}		
\usepackage{latexsym}		
\usepackage{bm}			
\usepackage[english]{babel}
\usepackage{color}		
\newcommand{\lsim}{\mbox{\rlap{\hbox{\lower2pt\hbox{\ensuremath{\sim}}}}\raise2pt\hbox{\ensuremath{<}}}}%
\newcommand{\gsim}{\mbox{\rlap{\hbox{\lower2pt\hbox{\ensuremath{\sim}}}}\raise2pt\hbox{\ensuremath{>}}}}%
\renewcommand{\gtrsim}	{\ensuremath{\gsim}}
\renewcommand{\lesssim}	{\ensuremath{\lsim}}
\renewcommand{\farcs}	{\mbox{\ensuremath{.\mkern-5mu^{\prime\prime}}}}%
\definecolor{red}{rgb}{1.0,0.0,0.0}%
\definecolor{blue}{rgb}{0.0,0.0,1.0}%
\definecolor{purple}{rgb}{0.5,0.0,0.8}\newcommand{\purple}{\color{purple}}%

\graphicspath{{./}{figs/}}

\newcommand{\etal}	{\hbox{et~al.}}%

\newcommand{\JWST}	{\emph{JWST}}%
\newcommand{\code}[1]	{\textsf{\small #1}}

\newlength{\txw}
\newlength{\txh}

{\relax}%
\newcommand{\DELETED}[1]{{\purple\rule{0.1pt}{10pt}}\relax}%

\newcommand{\changed}[1]{#1}{\relax}%

\begin{document}

\title{JWST's PEARLS: dust attenuation and gravitational lensing in the backlit-galaxy system  VV~191}

\author[0000-0002-6131-9539]{William C. Keel} 
\affiliation{Department of Physics and Astronomy, University of Alabama, Box 870324, Tuscaloosa, AL\,35404, USA}


\author[0000-0001-8156-6281]{Rogier A.~Windhorst} 
\affiliation{School of Earth \& Space Exploration, Arizona State University, Tempe, AZ\,85287-1404, USA}
\affiliation{Department of Physics, Arizona State University, Tempe, AZ\,85287-1504, USA}

\author[0000-0003-1268-5230]{Rolf A.~Jansen} 
\affiliation{School of Earth \& Space Exploration, Arizona State University, Tempe, AZ\,85287-1404, USA}

\author[0000-0003-3329-1337]{Seth H.~Cohen} 
\affiliation{School of Earth \& Space Exploration, Arizona State University, Tempe, AZ\,85287-1404, USA}

\author{Jake Summers} %
\affiliation{School of Earth \& Space Exploration, Arizona State University, Tempe, AZ\,85287-1404, USA}

\author[0000-0002-4884-6756]{Benne Holwerda}
\affiliation{Department of Physics and Astronomy, University of Louisville, Natural Science 102, 40292 KY Louisville, USA}

\author{Sarah T. Bradford} 
\affiliation{Department of Physics and Astronomy, University of Alabama, Box 870324, Tuscaloosa, AL\,35404, USA}

\author[0000-0002-5404-1372]{Clayton D. Robertson}
\affiliation{Department of Physics and Astronomy, University of Louisville, Natural Science 102, 40292 KY Louisville, USA}

\author[0000-0002-2012-4612]{Giovanni Ferrami} 
\affiliation{School of Physics, University of Melbourne, Parkville, VIC 3010, Australia}
\affiliation{ARC Centre of Excellence for All Sky Astrophysics in 3 Dimensions (ASTRO 3D), Australia}

\author[0000-0001-7956-9758]{Stuart Wyithe} 
\affiliation{School of Physics, University of Melbourne, Parkville, VIC 3010, Australia}
\affiliation{ARC Centre of Excellence for All Sky Astrophysics in 3 Dimensions (ASTRO 3D), Australia}

\author[0000-0001-7592-7714]{Haojing Yan} 
\affiliation{Department of Physics and Astronomy, University of Missouri, Columbia, MO\,65211, USA}

\author[0000-0003-1949-7638]{Christopher J.~Conselice} 
\affiliation{Jodrell Bank Centre for Astrophysics, University of Manchester, Oxford Road, Manchester, M13\,9PL, U.K.} 

\author[0000-0001-9491-7327]{Simon P.~Driver} 
\affiliation{International Centre for Radio Astronomy Research (ICRAR) and the
International Space Centre (ISC), The University of Western Australia, M468,
35 Stirling Highway, Crawley, WA\,6009, Australia}

\author{Aaron Robotham} %
\affiliation{International Centre for Radio Astronomy Research (ICRAR), The University of Western Australia, M468,
35 Stirling Highway, Crawley, WA\,6009, Australia}

\author[0000-0001-9440-8872]{Norman A.~Grogin} 
\affiliation{Space Telescope Science Institute, 3700 San Martin Drive, Baltimore, MD\,21218, USA}

\author[0000-0001-9262-9997]{Christopher N.A.~Willmer} 
\affiliation{Department of Astronomy\,/\,Steward Observatory, University of Arizona, 933 N.\ Cherry Ave., Tucson, AZ\,85721, USA}

\author[0000-0002-6610-2048]{Anton M.~Koekemoer} 
\affiliation{Space Telescope Science Institute, 3700 San Martin Drive, Baltimore, MD\,21218, USA}

\author[0000-0003-1625-8009]{Brenda L.~Frye} 
\affiliation{Department of Astronomy\,/\,Steward Observatory, University of Arizona, 933 N.\ Cherry Ave., Tucson, AZ\,85721, USA}

\author[0000-0001-6145-5090]{Nimish P.~Hathi} 
\affiliation{Space Telescope Science Institute, 3700 San Martin Drive, Baltimore, MD\,21218, USA}

\author[0000-0003-0894-1588]{Russell E.~Ryan, Jr.} 
\affiliation{Space Telescope Science Institute, 3700 San Martin Drive, Baltimore, MD\,21218, USA}

\author{Nor Pirzkal} 
\affiliation{Space Telescope Science Institute, 3700 San Martin Drive, Baltimore, MD\,21218, USA}

\author[0000-0001-6434-7845]{Madeline A.~Marshall} 
\affiliation{National Research Council of Canada, Herzberg Astronomy \& Astrophysics Research Centre, 5071 West Saanich Road, Victoria, BC V9E\,2E7, Canada}

\author[0000-0001-7410-7669]{Dan Coe} 
\affiliation{AURA for the European Space Agency (ESA), Space Telescope Science Institute, 3700 San Martin Drive, Baltimore,vvv MD\,21218, USA}

\author[0000-0001-9065-3926]{Jose M.~Diego} 
\affiliation{Instituto de Física de Cantabria,
Edificio Juan Jord\'a, Avenida de los Castros s/n,
E-39005 Santander, Cantabria, Spain}
\affiliation{Instituto de F\'isica de Cantabria (CSIC-UC). Avenida. Los 
Castros, s/n. 39005 Santander, Spain}

\author[0000-0002-5807-4411]{Thomas J. Broadhurst} 
\affiliation{Department of Physics, University of the Basque Country UPV/EHU, E-48080 Bilbao, Spain}
\affiliation{DIPC, Basque Country UPV/EHU, E-48080 San Sebastian, Spain}
\affiliation{Ikerbasque, Basque Foundation for Science, E-48011 Bilbao, Spain}

\author[0000-0001-7016-5220]{Michael J. Rutkowski}
\affiliation{Department of Physics and Astronomy, 
Minnesota State University, Mankato, Mankato, MN 56001}

\author[0000-0001-7092-9374]{Lifan Wang}
\affiliation{Department of Physics and Astronomy, Texas A\& M University, Mitchell Physics Building (MPHY)
4242 TAMU, College Station, TX 77843-4242}

\author[0000-0002-9895-5758]{S. P. Willner}
\affiliation{Center for Astrophysics \textbar\ Harvard \& Smithsonian,
60 Garden St., Cambridge, MA 02138, USA}

\author[0000-0003-4030-3455]{Andreea Petric} 
\affiliation{Space Telescope Science Institute, 3700 San Martin Drive,
Baltimore, MD 21210, USA}

\author[0000-0003-0202-0534]{Cheng Cheng} 
\affiliation{Chinese Academy of Sciences, National Astronomical Observatories,
CAS, Beijing 100101, China}

\author[0000-0002-0350-4488]{Adi Zitrin} 
\affiliation{Physics Department, Ben-Gurion University of the Negev, P.O. Box
653, Beer-Sheva 8410501, Israel}




\correspondingauthor{William Keel}
\email{wkeel@ua.edu}

\received[Submitted to AJ\ ]{}

\setwatermarkfontsize{1in}

\shortauthors{Keel \etal}
\shorttitle{Dust and lensing in VV~191}

\begin{abstract}
   We derive the spatial and wavelength behavior of dust attenuation in the multiple-armed spiral galaxy 
VV\,191b using backlighting by the superimposed elliptical system VV\,191a in a pair with an exceptionally 
favorable geometry for this measurement. Imaging using JWST) and 
HST spans the wavelength range 0.3-4.5\micron\ with high angular resolution, 
tracing the dust in detail from 0.6 to 1.5~\micron\ . Distinct dust lanes continue well beyond the 
bright spiral arms, and trace a complex web, with a very 
sharp radial cutoff near 1.7 Petrosian radii. We present attenuation profiles and coverage statistics 
in each band at radii 14-21 kpc. We derive the attenuation law with wavelength; the data both within and between the dust lanes 
clearly favor a stronger reddening behavior ($R = A_V/E_{B-V} \approx 2.0$ between 0.6 and 0.9~\micron, 
approaching unity by 1.5~\micron) than found for starbursts and star-forming regions of galaxies. 
Power-law extinction behavior $\propto \lambda^{-\beta}$ gives $\beta=2.1$ from 0.6-0.9~\micron. $R$ 
decreases at increasing wavelengths ($R\approx 1.1$ between 0.9 and
1.5~\micron), while $\beta$ steepens to 2.5. Mixing regions of different column density flattens the 
wavelength behavior, so these results suggest a different grain population than in our 
vicinity. The NIRCam images reveal a lens arc and counterimage from a background galaxy at 
$z \approx 1$, spanning $90^\circ$ azimuthally at 2.8\arcsec\ from the foreground elliptical galaxy 
nucleus, and an additional weakly-lensed galaxy. The lens model and imaging data give a mass/light 
ratio $M/L_B=7.6$ in solar units within the Einstein radius 2.0 kpc.
\end{abstract}


\section{Introduction}

Dust is a key ingredient in both observed and physical properties of galaxies. It obscures and scatters stellar light, and enables star formation by acting as a catalyst for molecular gas formation. It is the ingredient in the ISM that most dramatically changes how we perceive a galaxy's starlight, and the site of much of the astrochemistry. On broad scales, correction for the effects of dust is an important part of many techniques for estimating astronomical distances. 

The cumulative effects of grains are often traced via attenuation (combined extinction and scattering) from the
ultraviolet to near-infrared regimes, and thermal emission in the far-infrared. Interpretation of the
UV-optical-IR (UVOIR) attenuation signature is limited by the need to understand the local radiation field and three-dimensional mix of stars and dust, both of which are features of the many routines designed to model the observed spectral energy distributions (SEDs) of a few individual galaxies in great detail \citep[e.g.,][]{Holwerda12, De-Looze12b, Gentile15, Verstappen13, Hughes14, Allaert15, De-Geyter14, De-Geyter15, Mosenkov16, Mosenkov18} or populations with preset priors \citep[e.g.][]{ Driver18,Bellstedt20,Bellstedt21,Thorne21,Zou22}.
\changed{In contrast to the extinction-based techniques for probing dust effects, the use of} far-IR emission
is independent of the need for any background source, but becomes progressively less
sensitive for colder dust temperatures, requires knowledge of the emissivity behavior of
grains much smaller than the wavelengths in question, and is generally observed at large angular diffraction limits. 

\changed{Star-by-star observations within the Milky Way, and recently
elsewhere in the Local Group, have been used to measure wavelength dependence of extinction
(absorption plus scattering). A widely-used parametrization is the 
CCM curve \citep{CCM}, which is characterized by the ratio of total to selective extinction in the $V$ band, $R = A_V/E_{B-V}$\footnote{where $A_V$ is the extinction in magnitudes at $V$ and $E_{B-V}$ is the color excess - difference in extinctions - between $B$ and $V$ bands}, the strength of the 2175-\AA\  bump, and UV slope. In the near-infrared bands, a power-law
form in wavelength has been found useful at least from 0.5--5$\mu$m, $A_\lambda \propto \lambda^{-\beta}$ with estimates of $\beta$ ranging from 2.07 (\citep{WangChen} to 1.65 (using the summary data in Table 3 of 
\citealt{ZhangYuan}). It has long been clear that there is not a universal extinction law, since variations are seen between  Galactic environments as well as between our part of the Milky Way and, for example, the Magellanic Clouds. 

In the context of almost all observations of distant galaxies, the extinction law (reflecting the grain properties)
is inextricably mixed with geometric factors involving the relatave distributions of emitting sources and grains (\citealt{Calzetti2001}, \citealt{Chevallard2013}), so we
are properly deriving the attenuation behavior which combines both factors. Many studies have derived attenuation laws from the color
behavior of galaxy samples and radiative-transfer models, or from 
models for the spectral energy distributions (SEDs) of galaxies compared to their observed properties, sometimes with a luminosity constraint including far-infrared emission (e.g., \citealt{Salim2018}). The \cite{Calzetti1994} attenuation
law, derived for the inner regions of starburst galaxies, has proven to be of surprisingly wide applicability, a reasonable fit to the behavior of whole galaxies as well as the outer regions of some star-forming galaxies. However, its applicability into the infrared is limited because, as defined, this relation goes to zero at 2.2 $\mu$m.

Generically, attenuation laws for galaxies or large parts of galaxies will be flatter (have a smaller wavelength dependence) than the extinction laws applying to individual stars, due to the mixing of regions of differing extinction within the area measured (and possibly significant effects of scattering into the beam). This is true for the \cite{Calzetti1994} form versus the \cite{CCM} curve, and for the range of
near-IR indices $\beta=1.0-1.4$ found for integrated galaxy light by \cite{Salim2018} compared to the steeper values $\beta \approx 1.8$ found for single stars as noted above. Our HST and JWST data allow a new
approach to this issue in the $0.6-4.4$ $\mu$m range.

In earlier work,} we have explored the use of backlit (occulting) galaxy pairs to improve our understanding
of attenuation in galaxies. In these cases, a background galaxy backlights a
large part of a foreground system, providing spatially continuous light  (ideally from
a smooth E or S0 galaxy) so that all the dust structure can be detected. While limited
to a small (random) subset of galaxies, this technique preserves the angular resolution of
UVOIR imaging, \changed{avoids the bias of small background objects toward preferential detection in more transparent spatial ``windows \citep{Holwerda2007},} and works equally well independent of grain temperature (specifically even for the coldest grains, whose emissivity can be very low for far-IR and submillimeter detection). 

This technique has been applied to a range of spirals, with the sample size growing as survey data improved the potential sample size. 
Small samples in ground-based images were considered by \cite{Keel1983},
\cite{WK1992}, \cite{WKC2000}, 
and using HST data by \cite{Elmegreen2001},
\cite{KW2001a}, \cite{KW2001b}, and \cite{Holwerda2009}.
This technique was extended into the ultraviolet, mostly using GALEX data, by
\cite{Keel2014}, with more detailed treatment of the uncertainties introduced
by imperfect symmetry in the galaxies. Addition of spectroscopic information can improve the separation of contributions from the two galaxies \citep{Domingue00} and provide a finer wavelength sampling of the attenuation \citep{Holwerda13a}. When both far-infrared (FIR) and backlighting measurements are possible, \cite{Domingue99} found broad consistency in the dust masses derived in these complementary ways. These studies have provided estimates of
attenuation and dust masses, hints of fractal dust structure in well-resolved spiral arms,
and has shown that the \cite{Calzetti1994} effective attenuation law applies to dust across
many spiral disks as well as in the star-forming regions for which it was initially derived.
Comparison with simulations indicates that this behavior with wavelength, quite different from that
characteristic of extinction measured star-by-star in nearby galaxies, results from the mix of 
optical depths within each resolution element. \cite{KW2001a} showed, using HST data,
that the derived attenuation law becomes steeper as the resolution element becomes smaller,
across size ranges 50-200 pc, fitting with this idea.
A handful of galaxies are known to host dust lanes or patches so far from their centers that the grains must be very cold ($<15$ K) \citep{Holwerda2009}, which is in the right sense to account for a low-temperature component of the overall emission of galaxies
including {\it Herschel} data \citep[e.g.,][]{Bourne12,Ciesla12,Smith12a,Cortese12,Cortese14b,Clark15,Beeston18}. 


``Prime Extragalactic Areas for Reionization and Lensing Science'' (PEARLS) is a \JWST\ GTO program (PI: R.~Windhorst) designed to provide the community with medium-deep imaging in up to eight near-infrared filters \citep{PEARLSoverview}.  As one of the main program goals is to trace the evolution and assembly history of galaxies over cosmic time, the impact of extinction and scattering by interstellar dust grains on the observed tracers of such evolution poses a potential source of bias and uncertainty. 
To assess this issue, one of the PEARLS targets is a pair of galaxies with a fortuitous chance alignment on the sky: a face-on spiral galaxy partially backlit by an elliptical galaxy. In such a configuration, the properties of
the dust in the foreground spiral galaxy can be measured in an absolute sense, and to very low dust column densities, and in both arm and interarm regions. This target, VV\,191, serves as a GTO pilot program to demonstrate the power of \JWST\ for future GO surveys that target a larger and more diverse sample of galaxies. In addition, unknown to us when selecting the target and clearly revealed by the JWST data, one of the components of the galaxy system observed here strongly lenses a distant background galaxy, illustrating multiple configurations and outcomes of overlapping galaxy images at once.

Previous work on dust in backlit galaxies has concentrated on grand-design spirals, often
2-armed systems,
where application of symmetry to model the foreground galaxy is manageable. VV\,191 projects the
outer disk against such a bright part of the background galaxy that we can complement earlier work by studying the disk attenuation of a multiple-armed spiral, where the dust lanes have flocculent structure. VV\,191 was
initially recognized as an occulting, rather than merging, system during the Galaxy Zoo survey for such
pairs \citep{GZcatalog}. It nearly matches the schematic ideal for analysis of such a pair,
with a face-on spiral roughly half-projected against a background E or S0 galaxy (itself 
nearly circular in projection), and the spiral nearly face-on (photometric axial ratio within 0.15 of circular at all radii). The backlighting near the edge of the disk is so strong that uncertainties in correcting for the spiral's own light barely contribute to the error budget in retrieving the attenuation. The analysis
technique and uncertainty analysis are described in detail in section \ref{sec-analysis}.

We present here a combination of HST and JWST imaging studies for VV\,191, showing the power of carrying this technique into the near-infrared (NIR), using a galaxy pair with nearly ideal geometry for this analysis. Adding the near-IR bands was expected to provide more leverage for measuring the reddening slope, and give accurate attenuation measurements in regions that might be effectively saturated at much shorter wavelengths. As it happened, the F090W and F150W JWST data also provide higher spatial resolution of dust structures.

\section{Observations and data analysis}

\subsection{The galaxy pair VV 191 and its overall properties}

The VV atlas\footnote{English translation hosted at https://ned.ipac.caltech.edu/level5/VV\_Cat/frames.html}
\citep{VVcatalog} shows VV\,191b as the spiral to the northeast, and VV\,191a as the background elliptical 
galaxy. NED uses the same convention; we are observing attenuation by VV\,191b against the light of 
VV\,191a. Because existing values are compromised by blending, we redetermined the Petrosian radius (in the SDSS convention) for the spiral VV\,191b after subtracting a model for the elliptical galaxy and using only the non-backlit portion of the galaxy, giving 12.5\arcsec=12.5 kpc in the F606W filter (essentially $r$ band). This is a large galaxy, and we detect backlit dust beyond a radius 20 kpc (1.7 Petrosian radii). Virtually the entire radial range where we trace backlit dust (14-20 kpc) is outside the radius where even the NIRCam images show distinct spiral arms in starlight\changed{, making modeling of the spiral's light unusually tractable}; the interstellar medium still maintains a spiral pattern in this outer zone. Symmetry and continuity show that long, organized dust lanes occur just inside the stellar arms, but are not confined to these locations.

VV\,191 was originally selected as having such a large redshift difference that interaction between the components could be ruled out. SDSS DR7 gave a redshift for one component of $z=0.051$, while NED quoted one for the other galaxy of $z=0.037$. However, after the STARSMOG snapshot program observed it, by the release of SDSS DR15, SDSS measured the second galaxy and gave an accordant value (current values are $z=0.0513$ for the elliptical and $z=0.0514$ for the spiral). Evidence for whether these galaxies might be affected by mutual gravitational interaction must then be sought from the galaxy images themselves. \changed{Neither galaxy shows the kind of asymmetry which suggest interactions within radii $\approx20$ kc, indicating that they are at least the sum of these radii apart along the line of sight.} This redshift with the current consensus cosmology gives an angular scale of 1.008 kpc arcsecond$^{-1}$ \citep{WrightCalc}. The galaxy nuclei are 20\farcs 4 apart (20.6 kpc in projection). \changed{The FWHM resolution in our images projects to values from 60 pc (F090W) to 180 pc (F444W).}

\subsection{HST imaging}

An initial HST image of VV\,191 was obtained as part of the STARSMOG snapshot program (HST program 13695), which targeted potential backlit-galaxy systems from the Galaxy Zoo and GAMA samples. Using the Wide-Field Camera 3 (WFC3) with F606W filter, we obtained a total exposure time of 900 seconds, in two subexposures with a small dither motion between. This passband, centered near 6060 \AA\ , is broadly similar to SDSS $r$. 

\changed{Hoping to measure} the slope of the attenuation law over a broad wavelength range, we also obtained WFC3 images in the UV F336W and F225W bands under HST program 15106 (PI Holwerda). A single-orbit (2749-s) pair of dithered exposures was obtained in F336W, while three single-orbit exposures totalling 8874 s were carried out in F225W band. As expected, the signal levels from the red elliptical galaxy were low enough to make adequate treatment of cosmic-ray events a challenge. 
\changed{We were able to retrieve a transmission map with usable precision for a
small dust-lane region near the elliptical-galaxy core in F336W, while for F225W we did not achieve the requisite signal-to-noise ratio in view of the faintness of the 
background in the UV. F225W is very sensitive in detecting regions of relatively unreddened star formation, and we find a useful correlation with the dust emission seen in the
NIRCam LW channels.}

\noindent\begin{figure*}[ht]
\centerline{
\includegraphics[width=0.95\txw, angle=0]{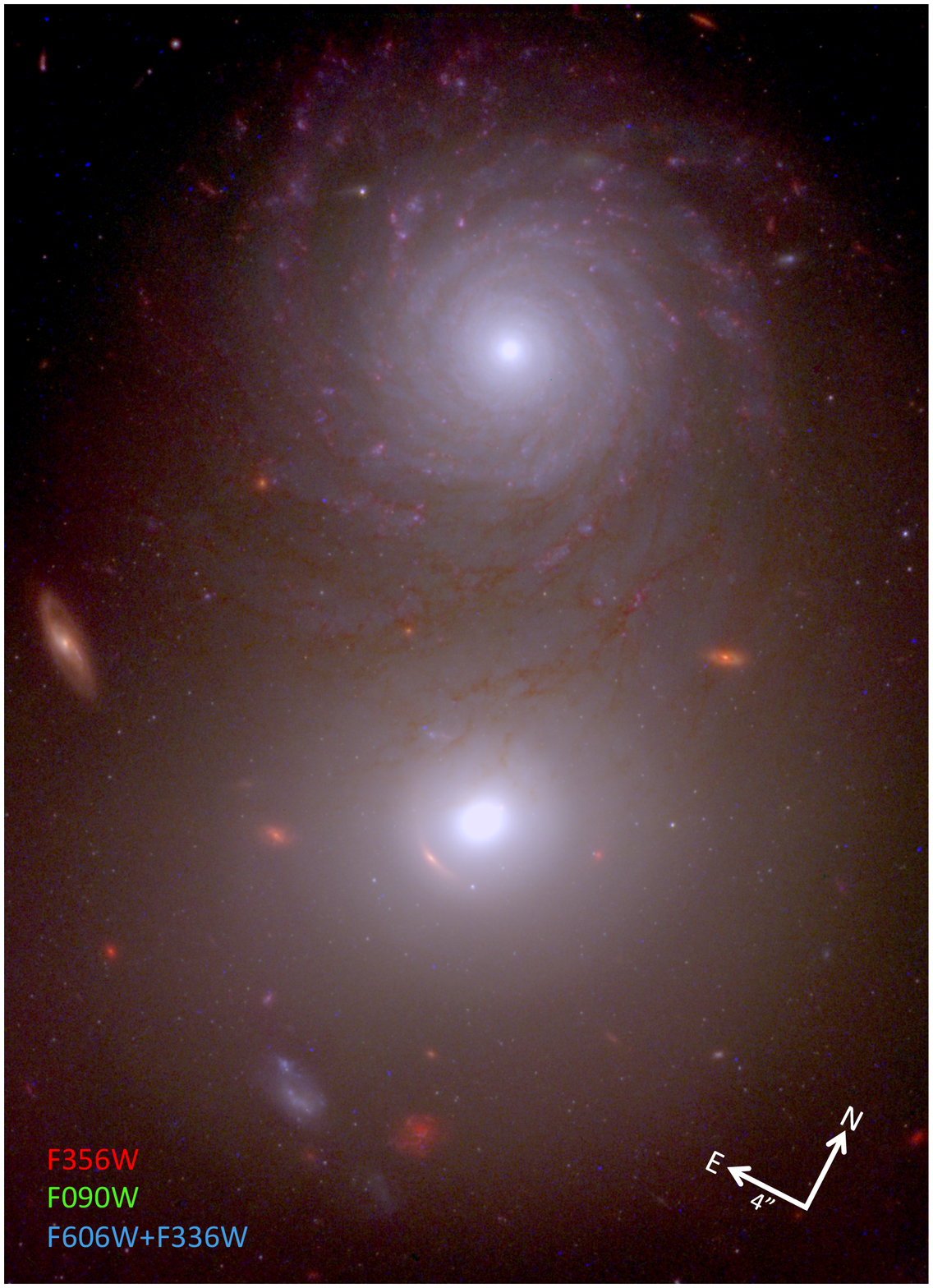} \ \
}
\caption{The overall geometry of VV\,191, illustrated with a composite of the HST F606W and F336W data in the blue channel, JWST NIRCam F090W in the green, and NIRCam F356W in red. Each is shown with a logarithmic intensity mapping, starting from slightly below the sky level to closely match SDSS-style sinh mapping. \changed{The field shown is $29.6 \times 55.0$\arcsec. The image is shown rotated by $27^\circ$ clockwise from cardinal orientation in this display, to align with the separation of the galaxies}. This shows the backlit dust lanes in front of the elliptical VV\,191a, as well as a probable \changed{gravitationally} lensed arc and counterimage behind VV\,191a, and additional background galaxies with a wide range of colors and likely redshifts.
\label{fig:vv191hstjwst}}

\end{figure*}

\subsection{JWST imaging}

VV\,191 was observed with NIRCam\footnote{http://ircamera.as.arizona.edu/nircam/in\_instrument\_overview.php} on 2 July 2021, \changed{quite early in the GO+GTO mission phase,} as part of the PEARLS GTO program (1176).
Both components of VV\,191 fit in a single short-wavelength (SW) detector quadrant, centered on the dust-overlap region, allowing freedom of
orientation and simpler reduction. Planned total exposures were 901 seconds in each of the
F090W/F356W and F150W/F444W filter pairs, all using MEDIUM8 readout with 3 groups and 3 dithers. The filters were chosen to combine high throughput with a long wavelength baseline; the 3.3~\micron\ PAH feature would appear within the F356W band, so we could in principle detect emission if strong enough \changed{(although we detect no such emission associated with the outer dust lanes)}. The NIRCam bands we observed are broadly comparable in wavelength to SDSS $z$, $H$, and the WISE [3.4] and [4.6] or Spitzer [3.6] and [4.5] passbands.


\bigskip

\subsection{Data Reduction}
\changed{
Reduction of the NIRCam data used pipeline version 1.8.0 and reference
files corresponding to CRDS context pmap-0995, as detailed by
\cite{PEARLSoverview}. This set of calibration
data improves flat-fielding and the uniformity of photometric zero points (especially between NIRCam detectors) compared to our initial look with pmap-0913, while the newer pipeline routines are much
better at rejecting `snowball" decay events which previously left circular artifacts. We paid particular attention to cross-checks of the flux scale (important only for the serendipitous lensing targets discussed below) and treatment of the ``1/$f$" readout noise, which manifests as gradual baseline drifts for each row along the $x$ direction \citep{Rauscher}. All detectors at each wavelength were drizzle-combined into single mosaic images, although our analysis here uses only the single short-wavelength (SW) detector encompassing the galaxies in VV\,191 and matching cropped regions on the long-wavelength detectors. Cosmic rays can be rejected almost perfectly with the multiple readout strategy  of the NIRCam detectors.}

The dust analysis relies only on differential measurements. For colors used in estimating photometric redshifts of newly-found lensed objects, \changed{especially since our initial analysis used an early version of the NIRCam photometric zero points, we used additional data for consistency checks on the magnitude scales.} We combined the SDSS photometry and spectrum of VV\,191a \changed{itself} with the known consistent behavior of the spectral energy distributions (SEDs) of giant elliptical galaxies.  We adopted the broadband data from \cite{Brown2014}, 
\changed{and 1.5-\micron\ spectrum from \cite{Francois2019}}. The scatter in broadband optical-IR colors for luminous ellipticals suggests that this procedure should be accurate to 5\% in flux, comparable to the initial precision of JWST absolute calibration \citep{Rigby2022}. \changed{In Table \ref{tbl-zeropoints}, 
we compare the derived magnitudes for the core 3\arcsec-diameter region of VV\,191a from NIRCam data with ``new" zero points (pmap-995) and corresponding values starting from SDSS and HST data, extrapolating redward with a composite SED as noted above. The scatter between the two sets of color indices is $\sigma=0.10$ magnitude, broadly consistent with the stated 0.05-magnitude accuracy of the NIRCam zero points and the scatter 
among elliptical galaxies in NIR colors ($\sigma=0.06-0.07$ magnitude depending on filter pair) from \cite{Brown2014}.
}


\begin{deluxetable*}{lcccl}
\tablecaption{\changed{Consistency color checks on VV\,191a nucleus}} 
\label{tbl-zeropoints}
\setlength{\tabcolsep}{7pt}
\tablehead{
\colhead{Color} & \colhead{AB (NIRCam)} & \colhead{AB (external)} & \colhead{NIRCam--external} & \colhead{Source}  \\}
\startdata
F606W-F090W  &  0.99 & 0.88	& 0.11	& SDSS spectrum\\
F090W-H		 & 0.57  & 0.62 &  -0.05  & \cite{Brown2014}, \cite{Francois2019} \\
H-F356W	 & 	-1.00 & 	-1.09	& 0.09	 & \cite{Brown2014} \\
F356W-F444W	& -0.43  & -0.51	& -0.08 & \cite{Brown2014} \\	
\enddata
\tablenote{F150W transformed to H using data in \cite{Francois2019}. F090W is consistent with SDSS $z$.}
\end{deluxetable*}

\bigskip

\subsection{Additional data processing}

\changed{We resampled the HST F225W, F336W, and (lower-resolution) F606W data to match the coordinate grid of the NIRCam drizzled mosaics, which were internally very well aligned through use of {\it Gaia} stars.}
In view of the change in how world coordinate system (WCS) information is included in the headers between HST and JWST data, we computed linear transformations between these images using sets of $\approx 10$ well-peaked objects in common. \changed{When needed to match point-spread functions (PSFs) between data sets}, we deal only with the Gaussian-like core. \changed{Residual effects of the diffraction spikes do appear in our attenuation maps at levels below 1\%, setting the precision floor in some regions near the core of the elliptical galaxy. The spikes normally appear at dynamic range of order 300 for the longer-wavelength bands,} and the Airy rings matter in the longer-wavelength bands only for dynamic range $>20$ (a level which our dust measurements reach only in the shorter-wavelength bands). 

\changed{The pipeline processing reduced the effects
of ``$1/f$" noise in our data, but the large angular size of the galaxies in VV\,191 meant that the most-used ways to use changes in sky level to track these variations along readout rows were not completely effective in the F090W and F150W data. We were able to substantially mitigate the residual ``$1/f$" effects in a heuristic way, by subtracting models for the large-scale structure of each galaxy, taking the median along readout rows with clipping of high and low values at $\pm 0.1$ MJy/sr, and subtracting this result from the data in each affected NIRCam filter after rotating to match the coordinate frame of the drizzled combination.} This success is due to the elliptical models incorporating essentially all galaxy structures which are not much smaller than the 251-pixel median window used to track the readout variations.

We illustrate the geometry of the VV\,191 system with a composite of the HST and JWST data sets in Fig. \ref{fig:vv191hstjwst}.
\bigskip

\subsection{Analysis technique}\label{sec-analysis}

Schematically, the fraction $T$ of transmitted light at each point in the foreground galaxy is 
derived from the observed intensity $I$ and model-based estimates of the foreground intensity $F$ and
background intensity $B$ (if observed by themselves, as judged from the non-overlapping parts of each galaxy) according to
$$ T={ {(I-F)} \over {B}}.  \eqno{(1)}$$
Transmission $T$ is closer to the observations than the more physically interesting optical 
depth $\tau$, and has better-behaved errors, so we start with $T$. To the extent that extinction-free sections of the background galaxy can be identified for the modeling, this technique provides an absolute zero point for attenuation.

When the galaxies in a backlighting system are close together, scattering may potentially alter the intensity and color of light coming through the foreground dusty disk. In this case, however, the extent of symmetric isophotes of the two galaxies indicates that they are not close enough for tidal influences to alter these (and have not done so for several crossing times). \cite{WKC2000} presented a calculation for a similar geometry and set of brightness profiles, showing that the differential scattering effects across the foreground disk (that is, the contribution which would not be subtracted by modeling the galaxy intensity profile) can be neglected as long as the galaxies are farther apart than the sum of their radii to the relevant region, almost certainly true in this case since the galaxies' isophotes are very elliptically symmetric (outside the overlapping regions) to still larger radii.
In addition, since scattered light is bluer than either attenuated or direct starlight, it would produce a trend in the color-attenuation relation with projected distance from background nucleus.

We modelled each galaxy using the procedure to fit ellipses to isophotes of
\cite{Jedrzejewski}, as implemented in the {\tt isophote} package of STSDAS.
The result is a smooth intensity model of each galaxy matching these isophotes
(whose centers may not necessarily coincide, although fixing the center may improve results when the arc of fitted data is short). This was done iteratively when needed, with progressively improved correction for each galaxy's effect on the other, both by subtracting the current model for the galaxy not being measured, and resulting improved masking of discrete features. This works very well for the
background early-type galaxy. It is obviously a \changed{cruder} approximation for the foreground spiral, usable in this case because the background light is so bright in comparison to the outer parts of the disk \changed{and the outer parts of the disk lack smooth stellar arms to further complicate modelling}. The form of Equation 1 shows how errors in each model affect the derived
transmission (point by point). Differences between the images and models gives typical (RMS) uncertainties
in regions of interest; there is not a straightforward way to show \changed{their effects} graphically when averaging spatially,  
because structure in the foreground galaxy has spatial correlations. The background galaxy in this system is demonstrably very smooth, leading to very small uncertainties where its light dominates over the foreground disk. The signal-to-noise ratio in the transmission image depends the brightness and symmetry of both galaxies, generally being best in the most strongly-backlit foreground region where there is significant dust attenuation. \changed{Pixel masks for fitting isophotes were produced for each image, iteratively improved by alternately subtracting the latest model for each galaxy before fitting the other. Including the lens arc, other background galaxies, star clusters, and foreground stars, typically 5\%} of pixels in the region of interest ($14^\circ$ on either side of the line between galaxy nuclei and radii 11-24.7\arcsec from the center of the spiral) were masked. 

\begin{figure*}[ht]
\centerline{
  \includegraphics[width=0.78\txw, angle=90]{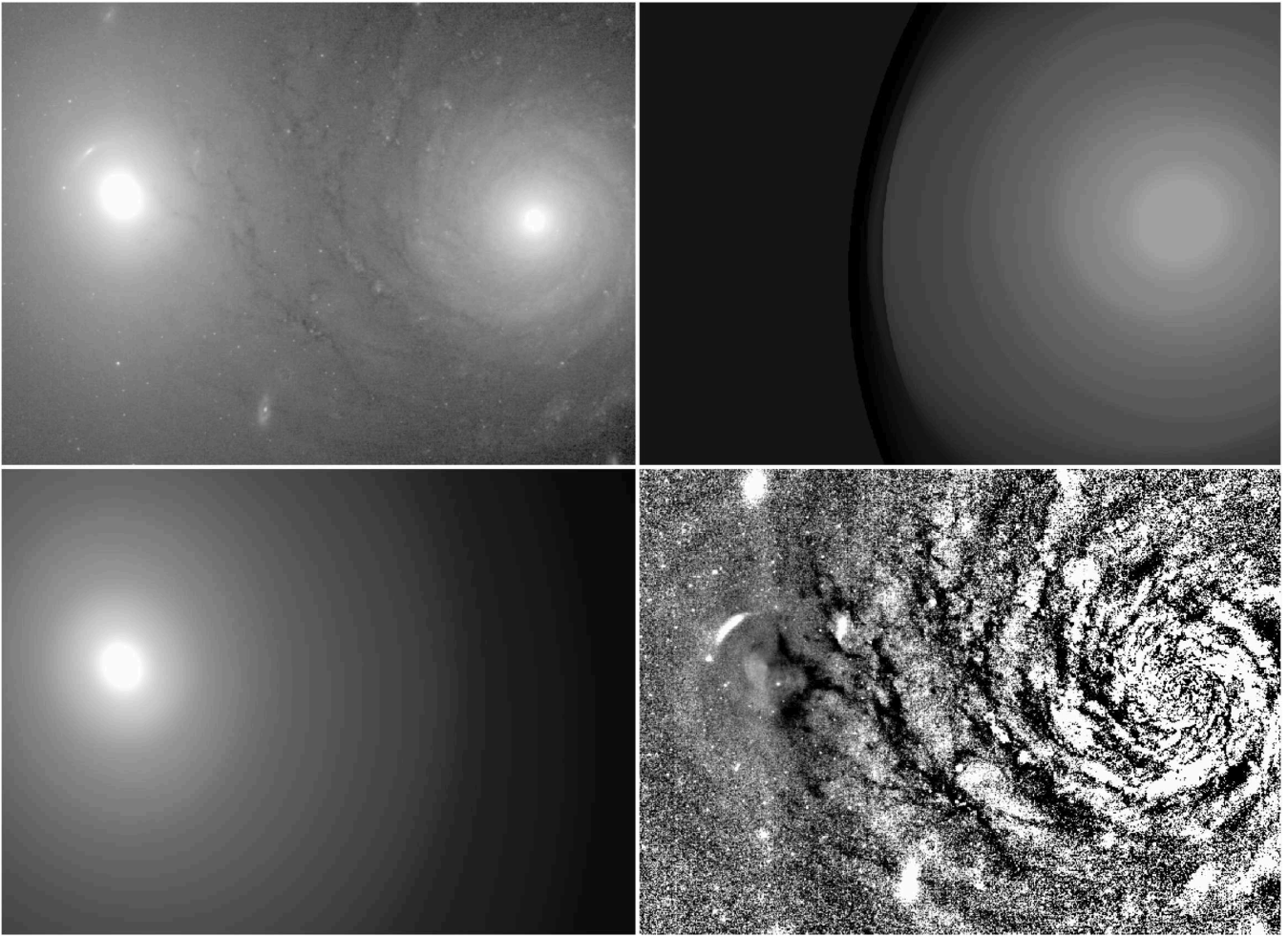}\ \ 
}
\caption{Steps in the dust analysis, shown using the NIRCam F090W data. Upper left, original image ($I$ in Eqn. 1). Upper right, smooth and symmetric model for the foreground spiral ($F$ in Eqn. 1) . Lower left, model for the background elliptical ($B$ in Eqn. 1). Lower right, derived transmission map for the foreground galaxy ($T = (I-F)/B$), which is meaningful only in regions where the errors due to galaxy structure are small 
\changed{(as shown in Section \ref{sec-spatial})}.
\label{fig:vv191analysis}}
\end{figure*}

Steps in this process are illustrated in Figure \ref{fig:vv191analysis} using the NIRCam F090W image. The longer-wavelength NIRCam images show speckled residuals in the brightest parts of the elliptical galaxy after model subtraction; we suspect this arises from image reconstruction with only 3 dither pointings, but this does not affect areas of interest for either dust or lensing studies. 

\noindent
\subsection{Uncertainty estimation}

Understanding the uncertainties in our attenuation maps informs our understanding of how far to analyze them. Since fine structure and the accuracy of modeling are more important contributions to the error budget than Poisson statistics or readout noise in some regions, we take an empirical approach to quantify the visual impression in our transmission (attenuation) maps. Starting with the model-subtracted data in each filter, we constructed a simple circular model of how each galaxy's structure, plus detection statistics, contributes to the overall uncertainties in the attenuation measurements. After subtracting our smooth model for each galaxy, we averaged standard-deviation values in annuli of each galaxy. For the half least affected by overlap, this standard deviation was evaluated in 15$^\circ$ sectors, omitting sectors including background galaxies, stars, or very bright star-forming regions, before deriving the average. For each galaxy, we fit a smooth empirical function of radius $r$ to these averaged standard deviations $\sigma$ 
using combinations of convenient basic functions (Gaussian, exponential, linear, truncated power law)
as interpolating devices, producing two-dimensional realizations of these function to compute the expected uncertainty per resolution element at each point in the attenuation maps. The signal-to-noise ratio of attenuation measurement reaches 20 (per $2 \times 2$-pixel resolution element) in two dark clouds in the F090W data. Throughout the region we analyze, the quality of the fit to the elliptical galaxy contributes more to the error budget than the mean foreground light for the outer disk of the spiral, allowing unusually well-constrained measurements of the effects of dust.

\changed{As a consistency check, we also evaluated the Gaussian scatter of derived attenuations
in ``blank" parts of each zone we analyze in Section \ref{sec-analysis}. Since resampling was involved in drizzle combination aligning HST to JWST data, and smoothing occurs when matching PSF core widths, we reduce pixel-to-pixel correlations via sparse sampling (every 3 pixels in each axis) for this exercise. The two approaches give results 
usually agreeing at the 30\% level in uncertainty for the NIRCam data, and at the factor-of-two level for the HST data}

We use these pixel-by-pixel uncertainty estimates in selecting regions for detailed analysis, and in fitting the slope of the reddening law between different filter bands. Deeper interpretation of these values will be limited by spatial correlations in the unmodelled galaxy structure, although for VV\,191 much of the backlit dust is in regions where the diffuse foreground spiral pattern is too faint to affect the pixel scatter measurably.


\section{Spatial structure of dust attenuation}\label{sec-spatial}

\begin{figure*}[ht]
\centerline{
  \includegraphics[width=0.75\txw, angle=90]{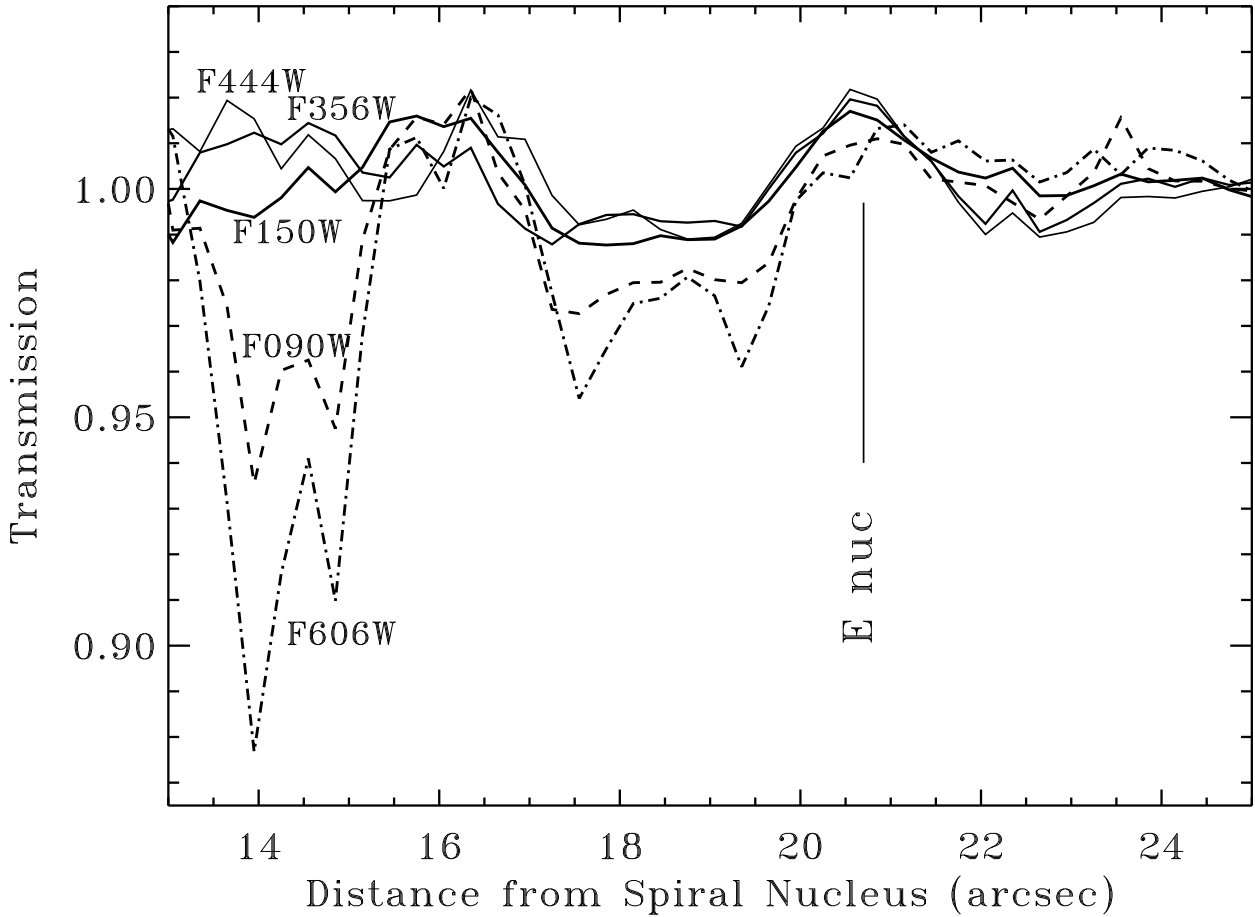}\ \ 
}
\caption{Azimuthally averaged transmission in F606W \changed{and all four NIRCam passbands}, from the zone of highest signal-to-noise ratio in the transmission map (13$^\circ$ of azimuth in the spiral disk, on either side of the line between spiral and elliptical nuclei). This quantifies the sharp outer edge to the dust lanes seen in Fig. \ref{fig:vv191analysis}, which is so complete that it is obvious even when averaged over all structure in the disk. \changed{Labelled curves show the various filter bands; F356W and F444W are distinguished by line thickness to reduce crowding. }
Unmasked pixels have been averaged in 10-pixel or 0.3\arcsec radial bins. The radial location of the background elliptical-galaxy nucleus is shown for reference. The standard deviation of the mean in each bin would be only slightly wider than the plotted lines, so the structure reflects real variation within the galaxy. With a $26^\circ$ sector width, spiral features give attenuation signatures where they cross it. At this redshift $z=0.051$, 1\arcsec = 1.008 kpc. \changed{Matching ripples in the profiles (most pronounced at radii 16.5 and 20.5\arcsec) for F150W, F356W, and F444W may indicate genuine asymmetric structure in one of the galaxies, since each image was independently masked and modeled. Their amplitude would
suggest an accuracy of 1.5\% from such structure. This motivates our fitting 
techniques for reddening behavior to incorporate such local systematic factors as nuisance parameters.}
\label{fig:vv191edgeplot}}
\end{figure*}

\begin{figure*}[ht]
\centerline{
  \includegraphics[ width=0.75\txw,angle=90]{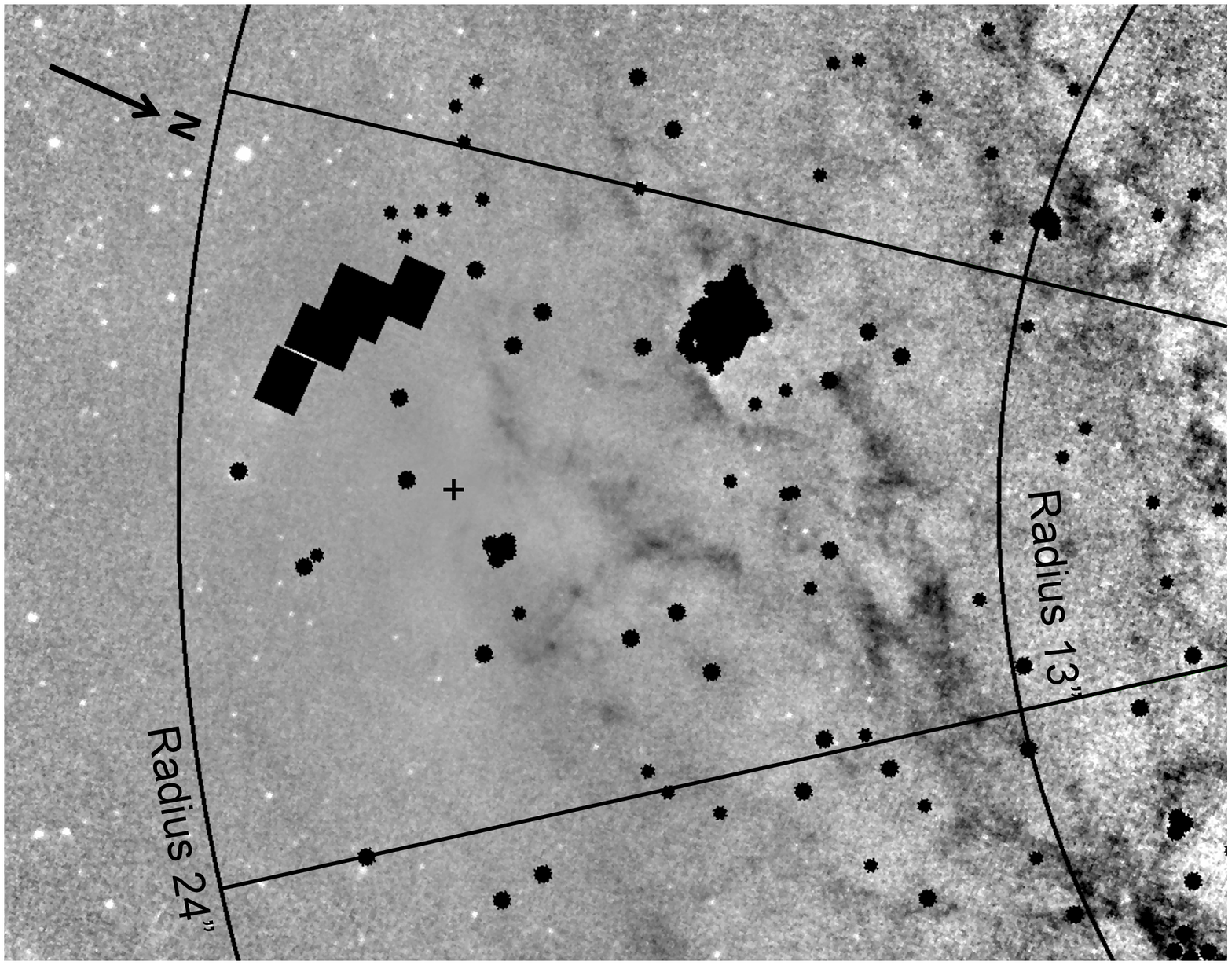}\ \ 
}
\caption{\changed{Sector of the disk of VV\,191b used for the measurements
in Fig. \ref{fig:vv191edgeplot}, overlaid on the F090W attenuation map. The orientation is the same as in Fig. \ref{fig:vv191analysis}, with north $24^\circ$ clockwise from the right. Circular segments show fiducial radii from the
nucleus of the spiral VV\,191b, while radial lines show the azimuthal boundaries of the
measured region with highest signal-to-noise ratio (center $\pm 13 ^\circ$). Black
areas indicate distant background galaxies, gravitationally lensed images, or foreground
star-forming regions masked from the fit here and in determining the wavelength
behavior of attenuation. The center of the elliptical galaxy VV\,191a is shown with a plus sign.}
\label{fig:zonemask}}
\end{figure*}

Important results came from the HST F606W snapshot image alone, informing
the selection of this pair for JWST observations. Most notable, the network
of dust lanes (both those tracing the spiral pattern and weaker lanes cross-cutting in a ```dust web")
has a sharp outer boundary, so pronounced that it appears in the azimuthally averaged
transmission profile (\citealt{Bradford}). \changed{This
appears in HST and JWST data (Figure \ref{fig:vv191edgeplot}). These profiles also show
dips in transmission where dust arms cross the sector of usable signal-to-noise. \changed{For reference, the region analyzed for the profiles in Fig. \ref{fig:vv191edgeplot} is outlined in Fig. \ref{fig:zonemask}.} The region more than 21.5\arcsec from the spiral core, with no evidence of dust attenuation, has a mean derived transmission within 0.4\% of unity in all five bands from F606W to F444W, furnishing a check on the accuracy and applicability of the galaxy models. However, there are ripples
in the ``zero-attenuation" level, including unphysical values greater than unity, at levels up to 1.5\% and closely matching
in bands from F150W to F444W. This amplitude is greater than plausible flat-fielding uncertainties in NIRCam\footnote{B. Sunnquist, private communication.}. We suspect these reflect mismatches between galaxy structure and our elliptically-symmetric models; this is an aspect of the low-level systematics which we attempt to deal with in fitting the reddening behavior in Section \ref{sec-reddening}. The fact that they are consistent between filters and confined in radius gives reason to think that the systematic errors are similarly confined in radius within the foreground disk so will be approximately constant across each of the regions we fit.}

\noindent
\begin{figure*}[ht]
\centerline{
  \includegraphics[ width=0.90\txw,angle=0]{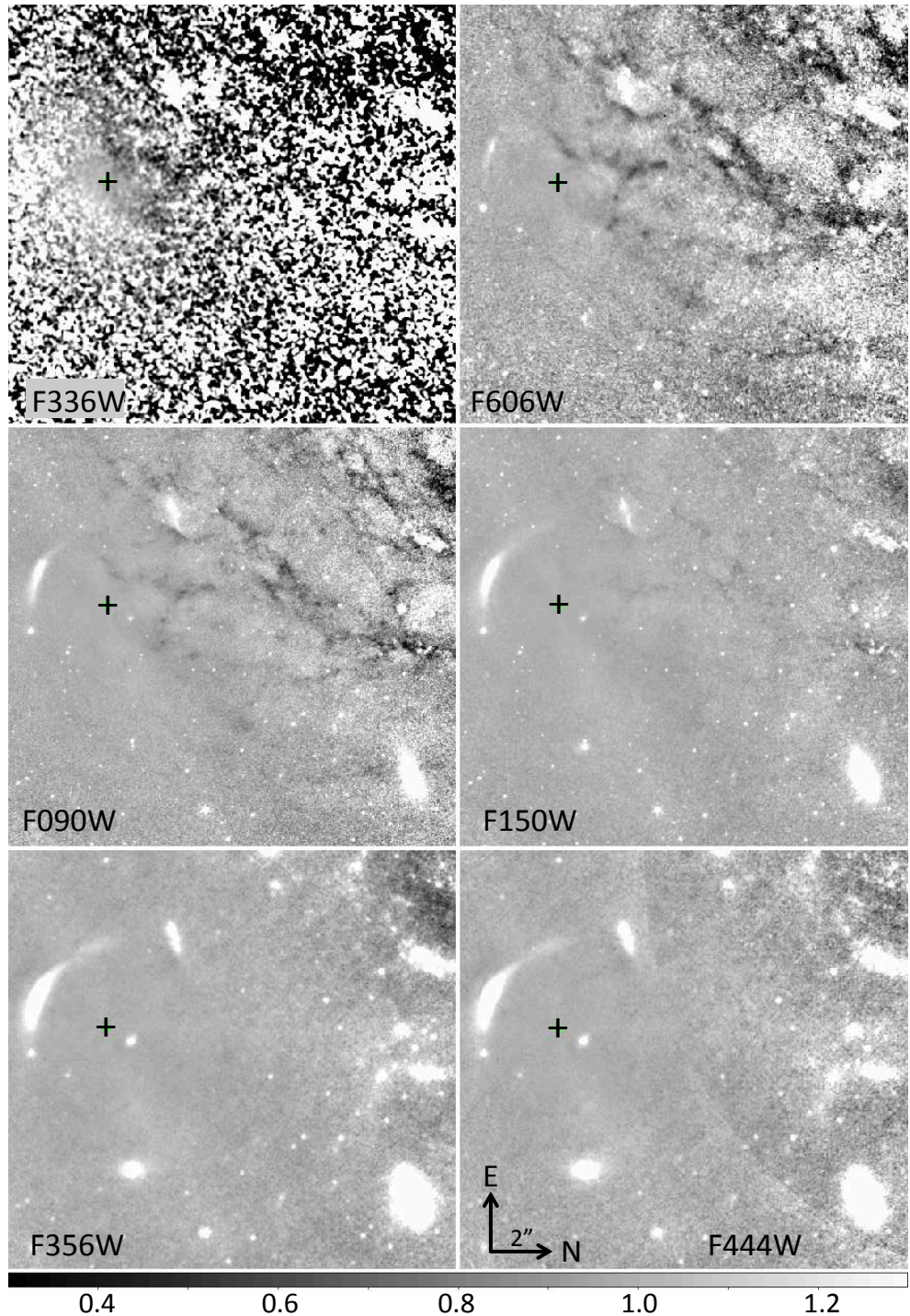}\ \ 
}
\caption{\changed{Transmission (and attenuation)} maps from HST and JWST data, spanning the observed range 0.33-4.4$\mu$m. These represent our estimates of the transmitted light at each point; regions with more dust, giving less transmitted light, appear darker. The edge of the dust disk almost exactly crosses the background nucleus (marked with a black plus sign). The combination of better galaxy symmetry and better S/N makes the dust features best detectable at 0.9\micron, despite the larger attenuation at shorter wavelengths. The background galaxy is so dim at 0.33$\mu$m that we can derive only spatially averaged quantities from that filter; for display, these data are median-smoothed by 0.15\arcsec\ here. All brightness scales are the same, \changed{from 0.3 to 1.3, as shown in the scale bar} (in transmission or residual intensity, to put unity at a conveniently visible level). The center of the spiral lies off the frame to the \changed{upper} right. Each image section spans $15.0 \times 15.9$\arcsec. Bright background objects and foreground stellar associations appear, which were masked for measurements of the dust effects. The F444W images shows ringlike artifacts to the right, from modeling the spiral disk, where masking during isophotal fits left poor residuals.
\label{fig:transmissionmaps}}
\end{figure*}
\begin{figure}[ht]
\centerline{
  \includegraphics[ width=0.50\txw,angle=0]{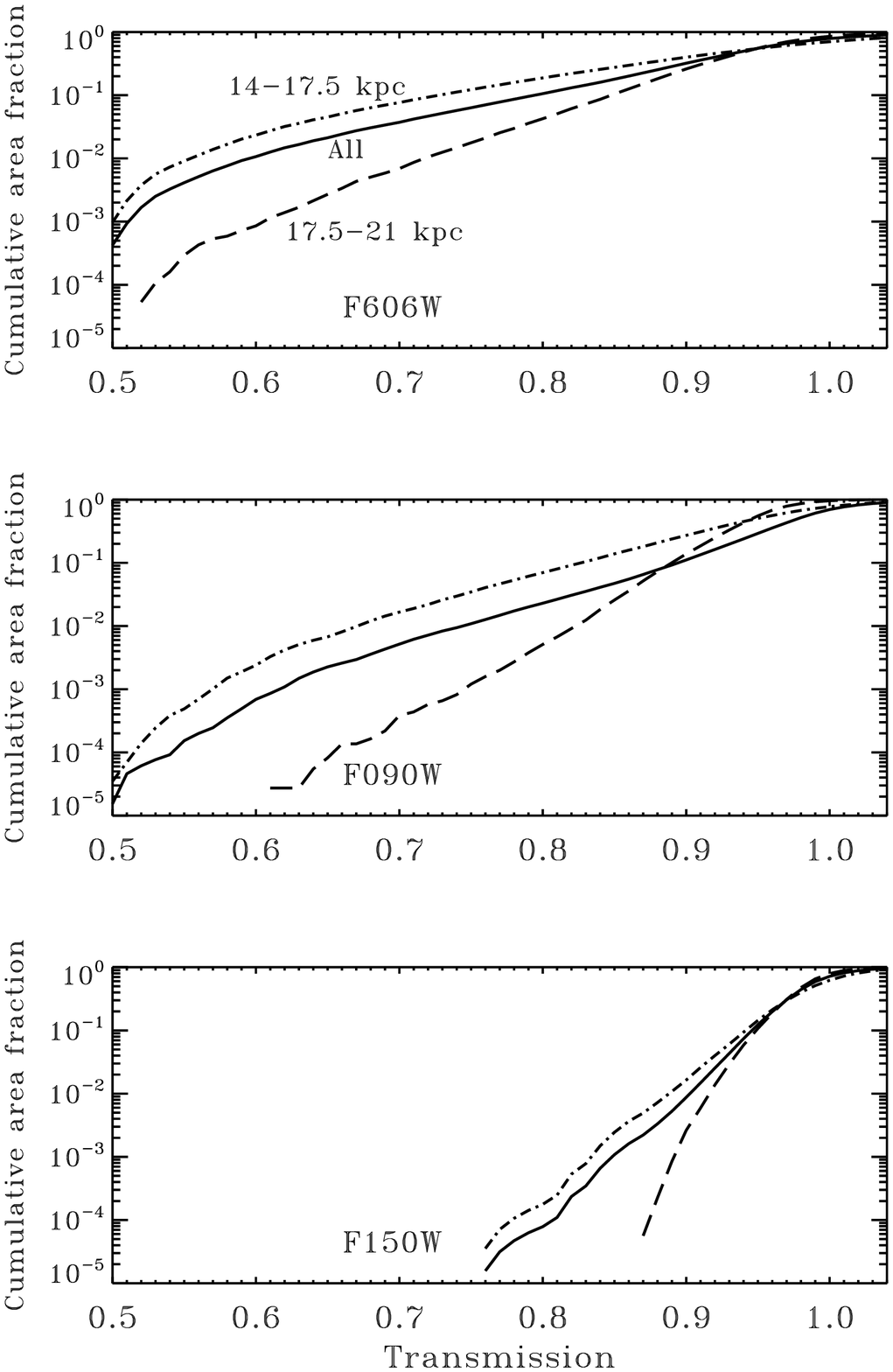}\ \ 
}
\caption{Cumulative distributions of area with transmission less than the value on the $x$ axis for the three best-measured filter bands \changed{within the best-measured disk sector as shown in Fig. \ref{fig:zonemask}}. The mapped radial range 14.5-21 kpc is shown as a whole (solid line) and broken into inner (dot-dashed) and outer (dashed) halves, as marked on the F606W panel. \changed{Pixels affected by unresolved sources,
background galaxies, and emission from star-forming regions have been masked.}
\label{fig:cumarea}}
\end{figure}

We illustrate these features across our observed wavelength range with the two-dimensional transmission maps from both HST and JWST data in Figure \ref{fig:transmissionmaps}. \changed{These illustrate the concentration of dust signatures to a system of lanes, some of which cut nearly perpendicular to the pitch angle of the overall spiral pattern (in what we term the dust web, some parts of which fit with spiral spurs of features as more often described). The shaping of these structures so far out in the disk appears to include processes
beyond the differential rotation which shapes so much of spiral patterns.

As a broad summary of the results, Figure \ref{fig:cumarea} shows} the cumulative distributions of transmission in broad radial bins. This is relevant to, for example, priors on color measurements of supernovae in galaxies of this type, and this technique offers the possibility of extension to a broader range of galaxy types and impact parameters. 
Figure \ref{fig:cumarea} provides a complementary view \changed{to the averaged radial profiles in Figure \ref{fig:vv191edgeplot}}, also useful in assessing the probability of a line of sight having a particular level of transmission all the way through the disk. This compares the cumulative (fractional) distribution of area with attenuation greater than a given level (transmission less than that value) for the three filters with the best dust measurements. This covers the well-measured range from 14.5-21 kpc, divided into inner and outer halves as well as showing the entire range. 

\noindent

\noindent

\noindent

\section{The broadband attenuation law in VV~191b}\label{sec-reddening}

The slope of the reddening-attenuation relation (reddening law) combines information on the grain populations and spatial distribution, when observed at the resolutions available for galaxies (in contrast to sight lines to individual stars, where scattering and blending of regions with different column densities are negligible).
We compare the distributions of points in the galaxy to the reddening trajectories predicted for various ratios of total-to-selective extinction $R = A_V/E_{B-V}$ or power-law index $\beta$.
For a range of $A_V$ values, we calculate the fraction of the flux from an elliptical galaxy emerging in each of our passbands, incorporating photon weighting \citep{Hogg2022} within the filter bands by folding in the SDSS spectrum of the core of VV\,191a, the 1.5-\micron\ spectrum of a typical luminous elliptical from \cite{Francois2019}, \changed{and 
in the near-UV F336 band, a composite of data for NGC 4472, incorporating averaged IUE spectra (\citealt{Norgaard}, \citealt{Oke}, \citealt{Burstein}) plus a Kitt Peak large-aperture spectrum extending near the atmospheric cutoff, all resampled to the redshift of VV\,191a.} Both photon weighting and finite passband widths mean that \changed{progressively} more flux is detected at \changed{increasing values of} $A_V$ than would be predicted \changed{from} the central wavelength alone.

 The linear resolution of these data ranges from 60--180 pc. Compared to the fractal-like distribution in the backlit spiral in AM~0500$-$620 \citep{KW2001a}, \changed{these values are in the regime} where we might expect significant ``graying" of the attenuation law as regions of different optical depth are mixed within each resolution element.

To the extent that a large part of the elliptical galaxy is unobscured so our model reflects its true brightness profile, this technique measures absolute attenuation, sensitive to gray attenuation as well as reddening.

\begin{figure*}[ht]
\centerline{
  \includegraphics[width=0.70\txw, angle=90]{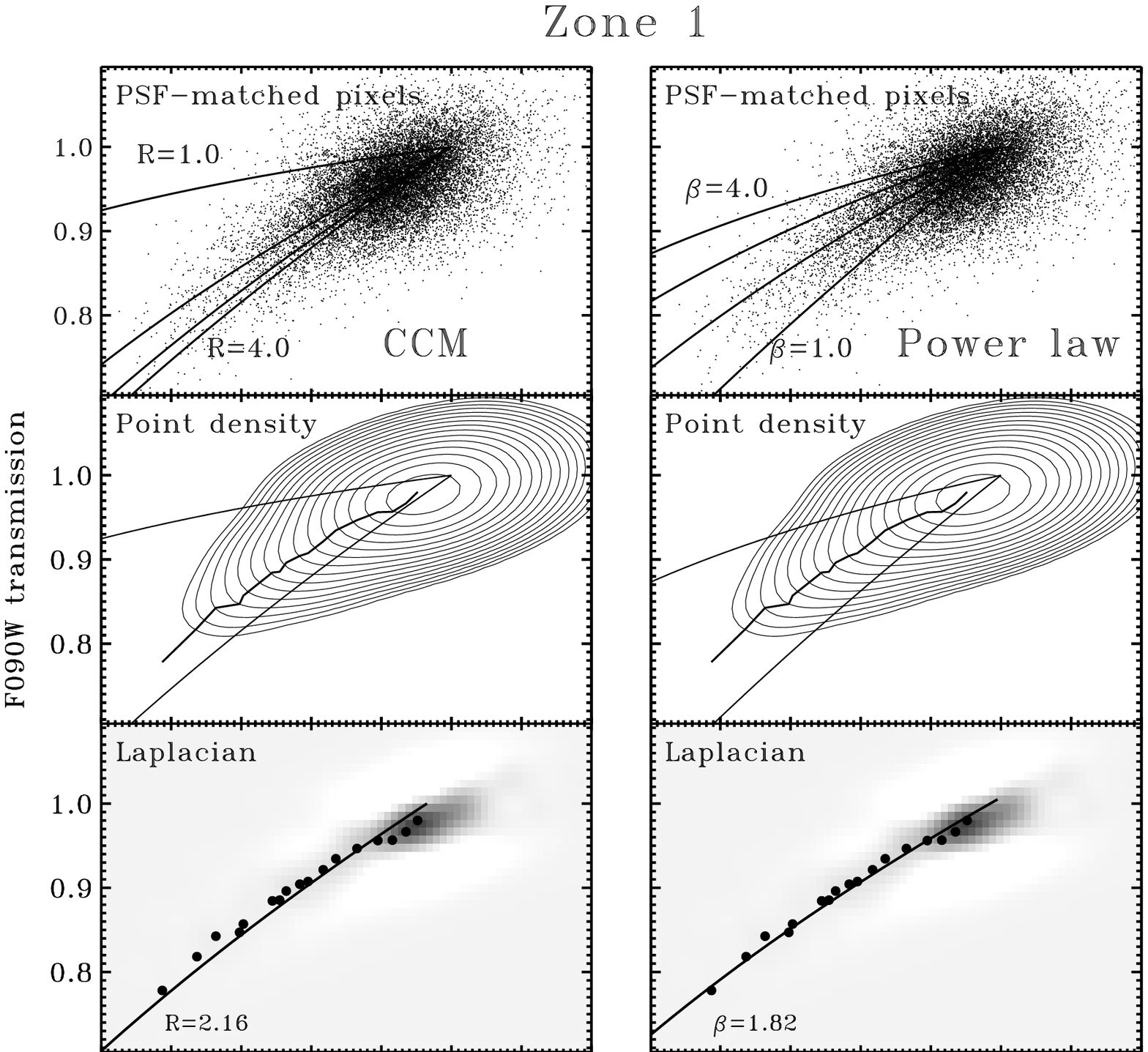}\ \ 
}
\vskip 0.5truein
\caption{Process for fitting reddening slope $R$ \changed{or $\beta$, comparing the distribution of points in transmission, after matching PSF core widths to the F606W value, in various pairs of filter bands with curves predicted by particular reddening forms}. Derived transmission values are shown in the \changed{F606W and F090W} filters, for all unmasked pixels in zone 1 of Figure \ref{fig:regions}. 
\changed{Curves in the left panels use the CCM reddening law with values of $R=1-4$ spaced by 1. Right panels use a power law in extinction $A_\lambda \propto \lambda^{-\beta}$ for $\beta=4$ to 1 in steps of 1. This zone emphasizes arm attenuation and therefore does not show the cloud of points near (1,1) indicating an area of no attenuation, although the cloud of most common values should still fall along the best-fit curve in the absence of systematic offsets. Gray attenuation would trace a straight line from the origin through (1,1), $\beta=0$ or an infinite value for $R$. The middle panels show logarithmically-spaced contours of the point distribution (after summing in bins of 0.01 on each axis), with thin lines marking the extreme curves from the top panels for reference. The bottom panels show this distribution in gray scale after filtering with a $3 \times 3$-bin Laplacian to trace its ridgeline, as a valley in the Laplacian gradient. The adopted fits to the ridgeline are shown as lines over the contour plots and as points over the Laplacian together with the best-fit curve for each parametrization. While a best fit can be found for either, each set of highly-constrained model curves shows systematic departures from the ridgeline of these data. }
\label{fig:howtofit}}
\end{figure*}

We measure the slope of the attenuation (reddening) law between each two of the three filters with the best attenuation measurements (F606W, F090W, F150W), matching the PSF core to the F606W data, by varying the slope between the filters \changed{(CCM parameter $R$ or power-law index $\beta$)} and fitting to the curve formed by the transmissions predicted for the two bands as the total extinction is varied (for bookkeeping and comparison, we use $A_V$, following the prescription of \cite{CCM} since the \cite{Calzetti1994} form is not defined in our redder filter bands). 
\changed{Fitting attenuation laws to these clouds of points poses several specific
challenges. The predicted transmission varying a given parameter ($R$ or $\beta$) for
each filter pair is a one-parameter family anchored at both (1,1) and (0,0), as shown in Fig. \ref{fig:howtofit}. In the regime of small attenuation (transmission $\approx 1$), even small
systematic errors in the data values \changed{(Section \ref{sec-spatial})} will exercise undue influence on the fitting results. \changed{Such small systematics would be multiplicative if due to a model shortfall for the elliptical galaxy or additive if due to structure in the spiral; they are always small in these data, less than 4\%, so we cannot distinguish and treat them as multiplicative for convenience.}
We explored several approaches to deriving these parameters, recognizing that
still more detailed modeling may be warranted, \changed{going beyond the scope of this analysis} and to be explored in future work.
Working in a low-attenuation regime into the near-IR bands, the fitting is
less well constrained than in, for example, the HST observations of NGC 3314
reported by  \cite{KW2001b}, where dust lanes with larger attenuations were
targeted including shorter wavelengths and only values relative to the diffuse surrounding
dust were retrieved, rather than the absolute attenuations we seek in this work.

These data are not accurate enough to discriminate between CCM and power-law forms
from 0.6-1.5 $\mu$m; there are values of each parameter which give virtually identical
attenuation behavior in each filter pair over the attenuation range we observe \changed{although the value of $A_V$ between equivalent parts of the two curves differs significantly at large attenuations).} 

\changed{Respecting the scatter of pixel values, which may be large compared to the changes we seek to measure, cases of asymmetric scatter in one or both filters, and the two-dimensional nature of the scatter since it may be comparable in each filter of a fitting pair, we experimented with
several ways to fit reddening slopes to data from each pair
of filters with significant dust measures over large regions (F606W, F090W, F150W). Ordinary $\chi^2$ minimization gives results very sensitive to small systematic offsets in the zero-attenuation level, since each fitting function by itself gives a one-parameter family anchored at both ends. We first tried fit to the curves bisecting the lowest-transmission sections of each point cloud, with a heuristic cut to reduce the influence of systematics at high transmissions, and also used binned versions of each point cloud to attempt fits following the ridgelines as defined by contour plots. Our final fits for the reddening behavior combine features of these approaches, with explicit allowance for systematics in galaxy removal. As well as the reddening slope (separately for $R$ and $\beta$, we fit for a multiplicative factor near unity (in practice $\pm 3\%$) for
each filter as nuisance parameters. In some regions, there may be an additional component of fine-scale structure in the foreground galaxy (that is, surface-brightness fluctuations). Inspection shows that actual noise and real structure (due to the distribution of luminous stars) contribute about equally in the long-wavelength F356W and F44W bands in the inner zones we fit, so the color of the brighter and dimmer spots depends on the stellar population. This would be manifested in the two-filter data by a component near unity transmission with a different slope than the larger-scale reddening behavior; we try to avoid this effect by fitting only
points with transmission $<0.98$ in the bluer passband of each pair.

Our fits start by summing the pixel values of transmission in a filter pair into a grid, normally summed in bin of 0.01 in each transmission value. }
Passing a $3 \times 3$-pixel
Laplacian gradient filter over such a smoothed array allows numerical retrieval of the
ridgeline of each distribution, where the attenuation parameter and systematic scaling
can be fit in a way analogous to $\chi^2$ minimization, minimizing the 2-dimensional deviations of each point on the gradient-derived ridgelines from the transmission curve, although the expected scatter is now
far removed from the individual pixel uncertainties. This approach therefore \changed{incorporates} the 2-dimensional nature of the uncertainties for each pixel, and gives a quantitative way of assessing the accuracy of the best-fit parameters. We present the results 
from this gradient-fit approach in in
Tables \ref{tbl-rfits} and \ref{tbl-betafits} for $R$ and $\beta$ respectively.  }

\begin{figure*}[ht]
\centerline{
  \includegraphics[width=0.78\txw, angle=90]{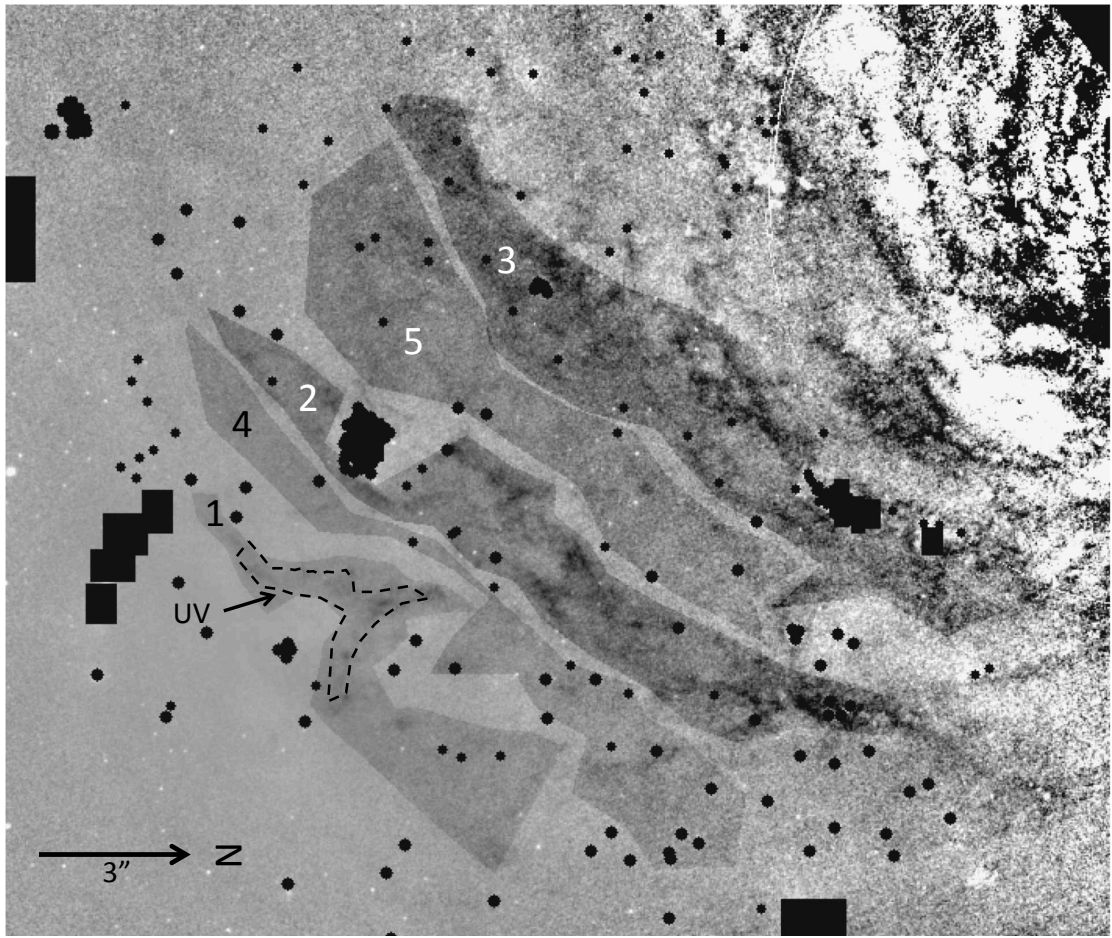}\ \ 
}
\caption{\changed{The F090W attenuation map with masked regions shown (black) and the five zones
chosen for fitting attenuation slopes in Table \ref{tbl-rfits}}. Left to right, they are regions 1, 4, 2, 5, and 3. Zones 1,2, and 3 are arm regions with organized dust lanes, while zones 4 and 5 are interarm regions with minimal coverage by distinct lanes. This subimage has \changed{north to the right, and covers an area $18.4 \times 21.7$\arcsec. Zones are marked by subtracting constant values for each from the attenuation map before display. The small area with a dashed outline labelled UV is the region used for F336W measurements, where the brightest background and large attenuation combine to give usable signal-to-noise ratios}.
\label{fig:regions}}
\end{figure*}

\begin{figure*}[ht]
\centerline{
\includegraphics[width=0.9\txw, angle=0]{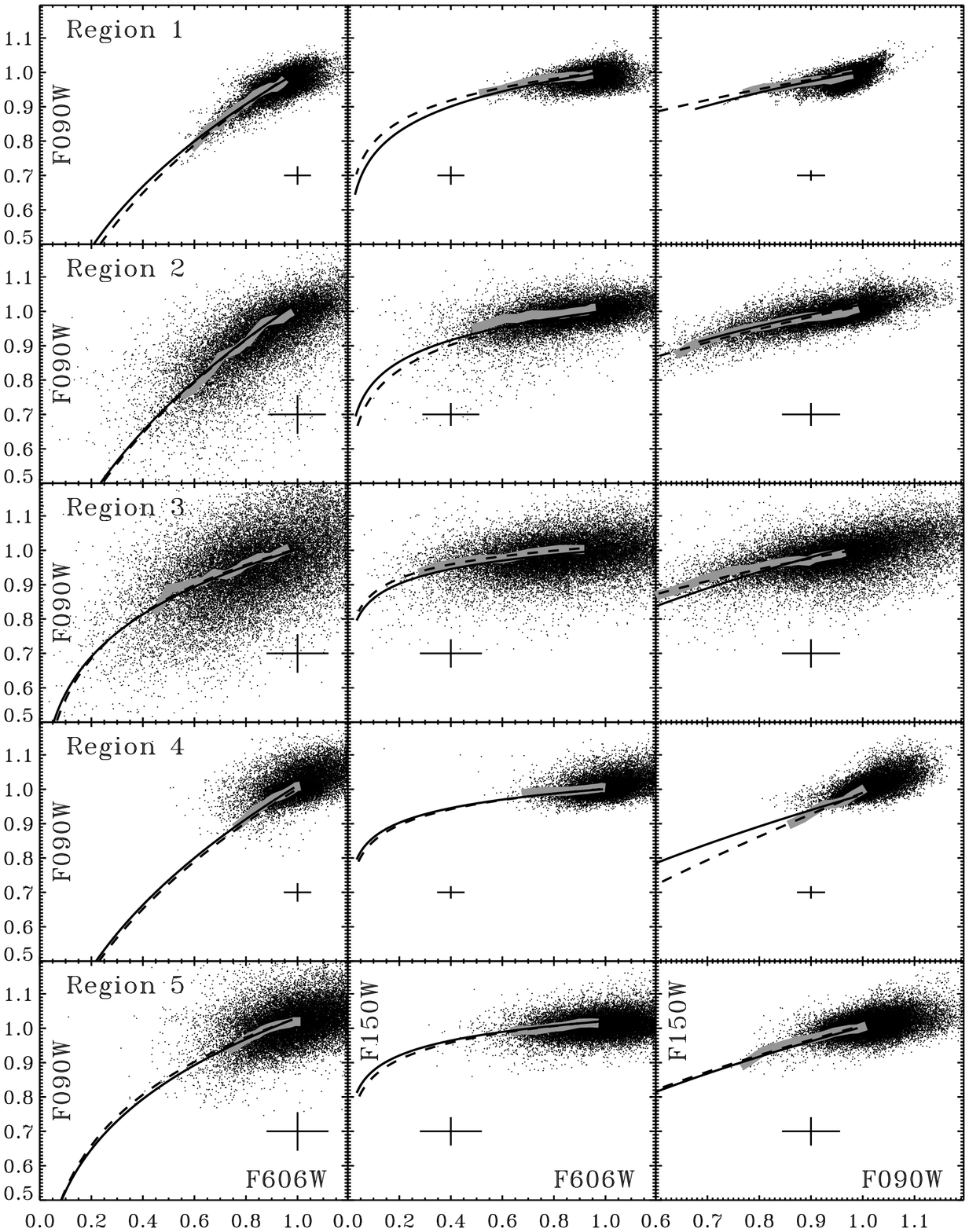} }
\caption{Point clouds for the zones in Fig. \ref{fig:regions}, with the trajectories corresponding to the adopted best-fit values of $R=A_V/E_{B-V}$ \changed{and power-law index $\beta$} as in Tables \ref{tbl-rfits} and \ref{tbl-betafits} superimposed. The derived transmission (residual intensity) in each pair of the filter bands F606W, F090W, F150W is shown pixel by pixel after matching PSF widths; the \changed{central} panels for F606W vs. F150W give the most leverage in fitting the \changed{overall} reddening law. \changed{Typical uncertainties per point are shown for each panel; since these are mostly function of distance from the two galaxy centers, they are nearly the same within each zone. The wide gray bands mark the ridgelines of the point clouds as traced with the Laplacian. The fitted slope values from Tables \ref{tbl-rfits} and \ref{tbl-betafits} are shown as curves, with the CCM values of $R$ shown as solid curves and the power-law slope results for $\beta$ as dashed curves.}
\label{fig:rvfit15}}
\end{figure*}

\changed{These fits have been} done for three regions spanning the arms as defined by dust lanes and two ``interarm" regions (Figure \ref{fig:regions}), which together \changed{largely} fill the region of highest S/N for attenuation. The fitting procedure is illustrated in Figure \ref{fig:howtofit}. \changed{We list the best-fit values for each region and
filter pair using each approach: values of $R$ in Table \ref{tbl-rfits} and 
power-law index $\beta$ in Table \ref{tbl-betafits}. The 2-band fits use pixel masks produced independently for each filter (and independent of the pixel masks used in fitting the galaxy models), to reject stellar associations in VV\,191b and background galaxies. When matching resolutions, the masked regions are propagated to include all pixels whose value would be affected by more than 0.05\% by a masked region in either filter band after the convolution, well below the intensity uncertainties in transmission values. The point clouds, ridgelines, and fitted 2-color transmission curves for
each pair of the three filters with widespread dust detection and each measured region are shown in Fig. \ref{fig:rvfit15}. This illustrates how similar the best-fit $R$ and $\beta$ curves are for each filter pair. The behavior for $R<1$, formally found for the F090-F150 filter pair, becomes highly nonlinear (and poorly determines from existing data), so we list $R=1$ as a minimum value and show this in Fig. \ref{fig:rvfit15}.}

\noindent

\changed{Across all regions and filter pairs, the value of $R = A_V/E_{B-V)}$ is very low, 1.94 from F606W-F090W and even smaller for filter pairs involving F150W (Table \ref{tbl-rfits}).This change with wavelength shows that the common fitting functions do not capture the broadband behavior in this situation (so a global fit across the well-measured bands from 0.6-1.5$\mu$m is not useful at this point).} 

\changed{We carried out analogous fitting for the index of a power-law attenuation law, where $A_\lambda = A_V (\lambda/\lambda_V) ^{-\beta}$, with results listed in Table \ref{tbl-betafits}. As with the CCM form, for each zone and filter pair, we did fits to the ridgeline of the point clouds as derived from the Laplacian transform of the gridded points, and as consistency tests, compare eyeball fits
both to the point cloud and to the ridgeline of the smoothed distribution of data points. We include weighted means for the entire data set and separately for the arm and interarm zones, derived by analysis of all points in each category rather than by averaging the zone values. For adequate signal-to-noise ratio in tracing the gradients, some zones (indicated with asterisks) had their gridded point distribution smoothed by a Gaussian of $\sigma=0.03$ in each transmission value rather than the default 0.02. Comparison of zones where both Gaussian values give useful results indicate that the scatter introduced by this choice is slightly larger than the listed uncertainty of the fit. Very broadly, regions where regions of different attenuation are blended within 
each resolution element will show flatter measured attenuation (large $R$, or $\beta$ closer to zero). The fitted values show trends of steeper slopes (more positive values of $\beta$) at longer wavelengths, smaller projected radii within the galaxy, and in interarm regions compared to the spiral arms.

Our fitting procedure gives useful bounds on the parameter values best fitting each data set. However, in most cases the 
attenuation tracks in these diagrams predicted by either CCM or power-law forms do not completely describe the data. Typically,
the data exhibit more curvature in the reddening tracks than the fits - that is, more reddening than predicted at small attention and less than predicted at large attenuation. This is broadly consistent with larger-attenuation pixels including areas with different extinction within each resolution element, but the situation invites further, more detailed modeling.}

\begin{deluxetable*}{lccccc}
\tablecaption{\changed{Gradient-Fit Results to Slope $R$ of the Attenuation Law}} 
\label{tbl-rfits}
\setlength{\tabcolsep}{7pt}
\tablehead{
\colhead{Region} & \colhead{Radius} & \colhead{Area} &  \colhead{F606W-F090W} & \colhead{F090W-F150W} & \colhead{F606W-F150W} \\
 & \colhead{(arcsec)} & \colhead{(pixels)} \\}
\startdata
1   & 19.4 &  15693  & $2.16 \pm 0.05$ &  $1.00 \pm 0.10^*$ &  $1.23 \pm 0.06$ \\
2   & 16.5 &  22013  &  $2.66 \pm 0.11$ &  $1.18 \pm 0.08^*$ & $1.14 \pm 0.06$ \\
3   & 10.5 & 28695  &  $1.20 \pm 0.05$ & $1.12 \pm 0.07^*$ & $1.10 \pm 0.04$\\
4   & 18.0 & 17168  &  $2.14 \pm 0.05$ & $  1.11 \pm 0.10^*$ & $  1.00 \pm 0.10$ \\
5   & 12.5 & 24734  &  $1.31 \pm 0.07$ & $  1.13 \pm 0.06^*$ & $  1.25 \pm 0.05$ \\
Mean & 14.6 & 107303 &  $1.94 \pm 0.06$ & $1.13 \pm 0.07$ & $1.10 \pm 0.06$ \\
Arms & 14.6 & 66401 & $2.07 \pm 0.05$ & $1.00 \pm 0.05$ & $1.13 \pm 0.04$ \\
Interarm & 147 & 41902 & $1.27 \pm 0.07 $ & $1.15 \pm 0.06$ & $1.00 \pm 0.04$ \\
\enddata	 
\tablenote{*Fitted with $\sigma=0.03$ smoothing of binned point distribution; otherwise, $\sigma=0.02$.}
\end{deluxetable*}

\begin{deluxetable*}{lccccc}
\tablecaption{\changed{Gradient-Fit Results to Power-law Slope $\beta$ of the Attenuation Law}} 
\label{tbl-betafits}
\setlength{\tabcolsep}{7pt}
\tablehead{
\colhead{Region} & \colhead{Radius} & \colhead{Area} &  \colhead{F606W-F090W} & \colhead{F090W-F150W} & \colhead{F606W-F150W} \\
 & \colhead{(arcsec)} & \colhead{(pixels)} \\}
\startdata
1   & 19.4 &  15693  & $1.82 \pm 0.08$ &  $2.90 \pm 0.10$* &  $2.67 \pm 0.09$ \\
2   & 16.5 &  22013  &  $1.68 \pm 0.10$ &  $2.53 \pm 0.07$ & $2.88 \pm 0.08$ \\
3   & 10.5 &  28695  &  $3.69 \pm 0.12$* & $2.68 \pm 0.10$ & $3.23 \pm 0.10$ \\
4   & 18.0 &  17168  &  $1.87 \pm 0.10$  & $1.0 \pm 0.2$* & $ 3.08 \pm 0.12$ \\
5   & 12.5 &  24734  &  $3.27 \pm 0.12 $* & $1.90 \pm 0.05$* & $3.00 \pm 0.10 $ \\
Mean & 14.6 & 107303 & $2.12 \pm 0.12$ & $2.48 \pm 0.12$ & $3.20 \pm 0.20$ \\
Arm & 14.6 & 66401 &  $2.00 \pm 0.10$ & $3.12 \pm 0.07$ & $ 3.04 \pm 0.06$ \\
Interarm & 14.7 & 41902 & $ 3.47 \pm 0.10 $  &  $ 2.04 \pm 0.09 $ & $3.54 \pm 0.08 $  \\
\enddata	 
\tablenote{*Fitted with $\sigma=0.03$ smoothing of binned point distribution; otherwise, $\sigma=0.02$.}
\end{deluxetable*}

\noindent

\noindent

The mean (all-region) values of $R$ (and $\beta$) in Tables \ref{tbl-rfits} and \ref{tbl-betafits} imply extinctions in these filter bands of $A_{F606W}/A_V = 0.83$, $A_{F090W}/A_V = 0.30$, and $A_{F150W}/A_V=0.047$. Slope fits to the longer-wavelength bands versus F606W give limits
(combining all 5 measured zones) $A_{F356W}/A_{F606W} < 0.035$ and
$A_{F444W}/A_{F606W} < 0.021$, which imply nonrestrictive limits on
the power-law index $\beta > 0.4$ and $\beta > 0.7$ respectively. Implicitly, the lack of relation between flux in these bands and attenuation measured at shorter wavelength shows the nondetection of emission from hot dust or PAH molecules in these (cold) dust lanes.

\begin{figure}[ht]
\centerline{
  \includegraphics[ width=0.38\txw,angle=90]{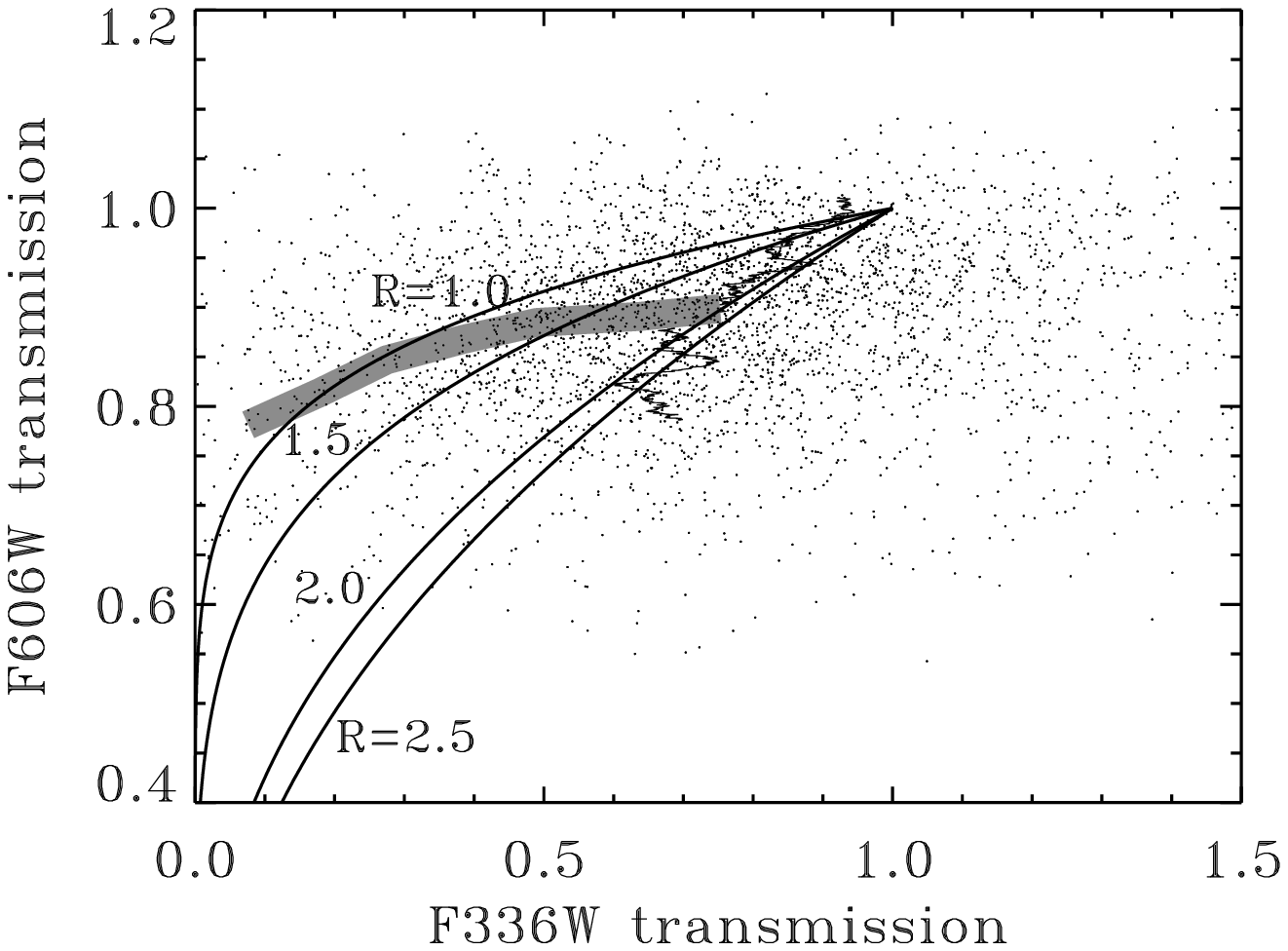}\ \ 
}
\caption{\changed{Fits to transmission measurements in the small region
used for F336W, comparing F336W and F606W transmission values. Individual pixels re show as points; the broad gray line traces the ridgeline of this point cloud using the Laplacian approach, from data in bins of 0.05 in 
each coordinate. For comparison, the jagged line shows a running mean over 199 points when sorted by F606W transmission, reflecting the much smaller error in F606W. The curves show the results of CCM models with the indicated $R$ values; we claim only that the data are consistent with $R=1.7\pm0.4$ and
strongly limited by the signal-to-noise ratio in F336W.}
\label{fig:f336atten}}
\end{figure}

\changed{Because the red color
of the background galaxy limits the ultraviolet S/N so severely, we use only a subset
of 3510 pixels in fitting region 1 for F336W encompassing the Y-shaped dust lane projected closest to the nucleus of VV\,191a, and constructed the 2D grid for fitting
the ridgeline of the F336W-F606W attenuation distribution using bin sizes of 0.05 in each filter. As shown in Figure \ref{fig:f336atten}, the extinction ratio between F336W and F606W bands
is best described by $R=1.7 \pm 0.4$ whether one fits the ridgeline, a running mean when sorted by F606W transmission, or by bisecting the cloud of individual pixel values. However, the gradient measures of the estimated ridgeline value show evidence of a further decrease at larger attenuations, more characteristic of a screen than the more likely
mix of extinction values. The low values of $R$ in some of our
fits may need to be understood as indicative, because the
behavior becomes so nonlinear for such low $R$ and there is little data on reddening behavior which is this strong.}

\noindent

The range of reddening behavior in galaxies as functions of galaxy type, metallicity, local environment, and averaging scale-size is surely broad and not yet understood. The most comparable previous results are from backlit galaxies observed with HST at linear resolutions $\approx 100$ pc. For disk dust in fairly luminous spiral disks, some systems show Milky Way-like reddening. The broad dusty arm in AM~1316$-$241 \citep{KW2001a} gives $R_V=3.4 \pm 0.2$, and multiple arm segments in NGC 3314 \citep{KW2001b} yield $R_V=3.5 \pm 0.3$ with no radial trend. However, the spiral arms in AM~0500$-$620 \citep{KW2001a} have $R_V=2.5 \pm 0.4$, while the foreground system in NGC 1275 has a very different, almost grey extinction law ($R>4$, formally much greater, from \citealt{KW2001a}). This galaxy is moving at high velocity through the intracluster medium of the Perseus cluster, so its grains may have been exposed to alteration by the hot gas. Comparison of spectral slopes in star-forming regions led \cite{Calzetti1994} to propose a reddening law which has seen wide application, giving consistent results not only for the starburst regions where it was first derived, but for the outer parts of spirals in the UV \citep{Keel2014}. Further work \citep{Calzetti2000} found a value $R_V=4.05$ for starburst regions in anchoring the parameters of this relation, which certainly responds to both the grain properties and the relative distributions of stars and dust in these intense star-forming regions. 

These kinds of measurements, summing over substantial regions, unavoidably mix regions of various optical depths, so this mix affects the results along with the (astrophysically more interesting) properties of the grain populations. \changed{We see the effects of this mixing by fitting power-law reddening slopes $\beta$ after block-averaging the transmission maps by factors 2, 4, and 8, propagating the masked regions in each case cross the 5 zones shown in Figure \ref{fig:regions}). In the mean, $\beta$, measured between F606 and F090W, decreases by 0.41 between 2 and 8-pixel averaging
steps, and decreases by a mean of 0.36 between F606W and F150W. Decreasing $\beta$ moves in the direction of grayer behavior, as expected for mixing regions of different attenuation. As the number of points decreases in this averaging, our ridgeline procedure ceases to be useful, so the block-averaged data were fitted using minimization about the predicted curves for the points themselves. Thus these are strictly comparable to each other but not necessarily to the ridgeline fits used for the full-resolution maps.}

The extinction behavior of the grains themselves is best derived when we can study their effects star by star, along essentially a single line of sight. This is how the classical Milky Way extinction curve was determined (albeit from a more limited part of the Galaxy than one might like), and has been done in the Magellanic Clouds and M31 \citep{Clayton2015} as well; the reddening law in the SMC in particular becomes much steeper than in the Milky Way going into the ultraviolet. Farther afield, supernovae (especially when observed polarimetrically) offer ways to isolate the effects of the grain properties, often implying $R\approx 2$ \citep{WangReview,SNpol}. There is clearly wide scope to improve our understanding of how the properties and distribution of grains vary within and among galaxies. \changed{Our results for VV\,191b, probing the outskirts of disk dust and extending to wavelengths where this behavior has been poorly known, show stronger reddening (steeper attenuation law with wavelength) than otherwise seen in galaxies, even with a possible role for mixtures of attenuation below the resolution limit flattening the observed behavior. This might hint at a very different population of grains, more weighted to smaller grains than in our neighborhood.}

\section{Gravitational Lensing}

Our first inspection of the NIRCam images revealed an object 2.8\arcsec from the nucleus of the elliptical galaxy VV\,191a, with a bright core and outskirts forming an arc nearly centered on the foreground galaxy. It is much redder than the elliptical galaxy, ruling out any sort of disrupted dwarf satellite and suggesting a gravitationally lensed image of a background galaxy. In this case VV\,191 is now seen to show both dust backlighting and lensing (retroactively earning its place in a project whose overall acronym includes ``Lensing").

\begin{figure*}[ht]
\centerline{
  \includegraphics[ width=0.85\txw,angle=90]{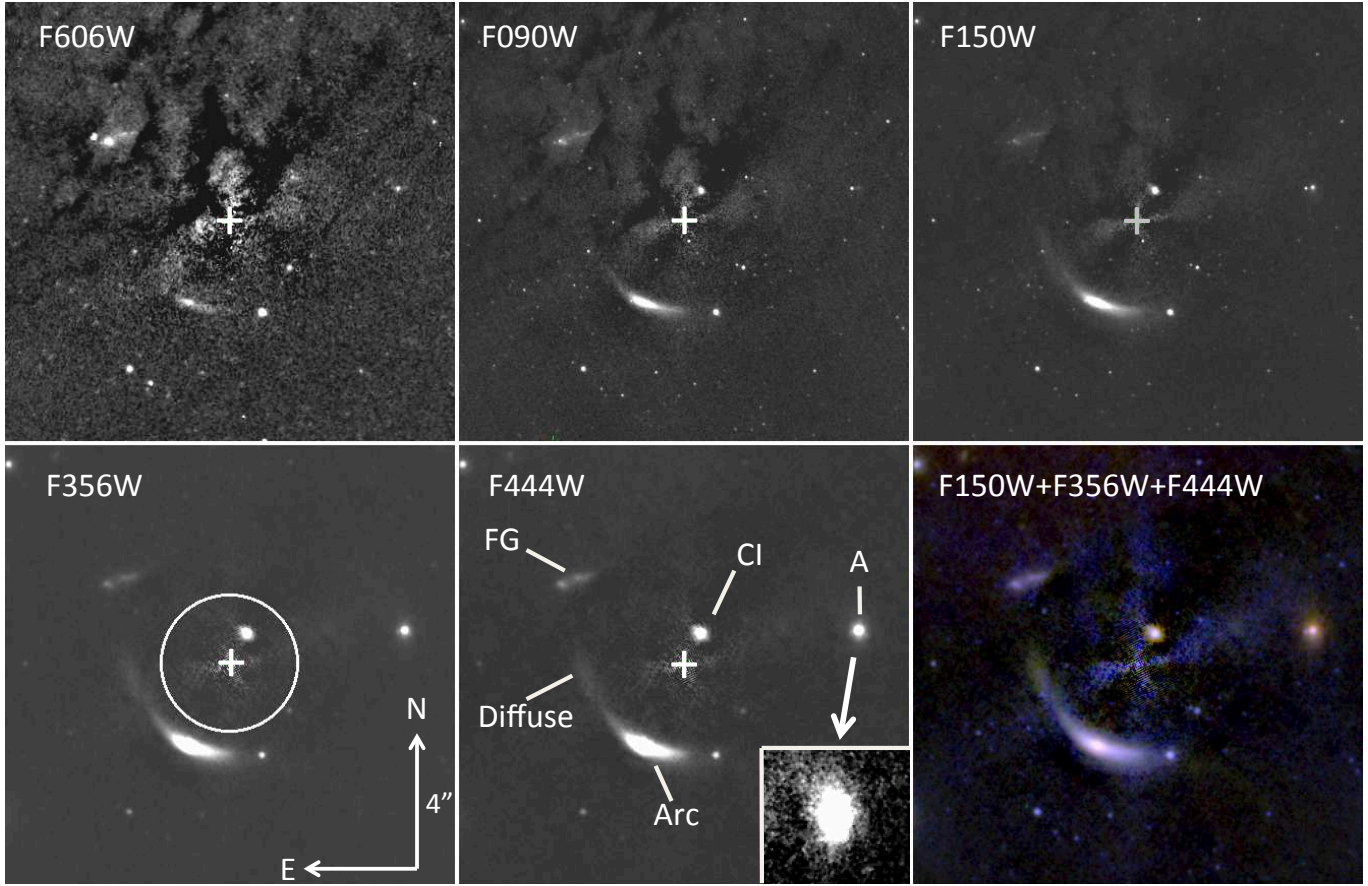}\ \ 
}
\caption{Lensed arc \changed{and counterimage}, shown after subtraction of the model for the foreground elliptical galaxy, with intensity scaling as appropriate for each. The top row shows HST WFC3 F606W and JWST NIRCam F090W and F150W bands, with NIRCam F356W and F444W at the bottom.   The nucleus of the elliptical galaxy VV\,191a is marked with a plus sign on the single-filter images. \changed{Other features discussed in the text or in Table \ref{tbl-arcphotometry} are labelled on the F444W image: the arc, counterimage CI, diffuse feature, and weakly lensed galaxy A. FG indicates an outlying star-forming region in the foregrounf galaxy.} At lower right is a logarithmically-scaled composite of F150W (blue), F356W (green), and F444W (red) data to show color structure in the arc and relative colors of arc, counterimage, and weakly lensed galaxy A (which is shown at higher contrast and magnified by a factor 1.5 in the inset for F444W). The arc has radius 2.83 \arcsec. \changed{The circle on the F356W image shows the modelled Einstein radius 2.01\arcsec. }
Model subtraction leaves signatures of dust from the outer disk of VV\,191b visible to the right of the elliptical nucleus, along with sub-percent residuals from fitting the elliptical-galaxy model near the core \changed{and minor contributions from the hexagonal PSF structure}. At the shortest wavelengths, foreground dust lanes cross the right side of the arc and affect its observed spatial profile.
\label{fig:arc6plex}}
\end{figure*}

This object is shown in Figure \ref{fig:arc6plex} after subtracting the model for the foreground elliptical as described in section \ref{sec-analysis}. It is so red that only its core was obvious in our HST F606W exposure. Galaxy-galaxy lens arcs often show two images, and indeed a \changed{similarly red}, slightly resolved object appears on the opposite side of the foreground nucleus from the arc, at projected radius 1.03\arcsec\ . We quantify the colors using aperture photometry (with radius 0.5\arcsec) for the brightest part of the arc and for the foreground nucleus, \changed{flux integrated within a segment of an annulus for the entire arc, and integrated Gaussian fits for the counterimage}. As we used data from an early version of the calibration pipeline while the photometric calibration was still in flux, we confirmed our understanding of the flux scale using the core of the elliptical galaxy VV\,191a and the consistent SEDs of luminous elliptical galaxies \changed{(Table \ref{tbl-zeropoints})}.
The possible counterimage is so small that we measure its count rates (WFC3) or signal in mJy (NIRCam) by fitting two-dimensional Gaussian profiles using IRAF {\it imexamine}. These results are shown in Table \ref{tbl-arcphotometry}, with ratios shown in magnitudes as well for quick interpretation. The listed errors include fluctuations in the residual background after model subtraction, which often exceed those from photon statistics.  

\changed{We estimate photometric redshifts for the lensed components based on the magnitudes in Table \ref{tbl-arcphotometry}, using the EAZY code \citep{EAZY}.
The arc has $z_{phot}=0.90 \pm 0.07$ (60\% probability) with secondary solutions
$z_{phot}=0.86 \pm 0.07$ (26\%) and $z_{phot}=1.17 \pm 0.08$ (14\%). Our initial analysis derived much larger values; the updated NIRCam zero points,
reflecting greater sensitivity in the longer-wavelength bands and especially F444W than
predicted, give a blue spectral slope between F356W and F444W which matches the long
side of the generic 1.6$\mu$m peak in galaxy SEDs. The 4000-\AA\  break is now
seen to fall between F606W and F090W bands. The filter selection, driven by the dust 
analysis, leaves a large gap between F150W and F356W which drives the large remaining
uncertainty in photometric redshifts. The counterimage \changed{has colors} consistent with
the same redshift, albeit with larger uncertainties. This modest redshift
fits with the small angular sizes of the lensed components, since the derived
redshift range is near the maximum in angular-size distance; at $z=0.9$ the
angular scale in the consensus cosmology is already within 10\% of its peak value at $z=1.6$.}

\noindent
We reproduced the image configuration, including the counterimage, with a two-component model for the lens: 
a singular isothermal ellipsoid (SIE) and an external shear. An initial optimization used the image centroids and flux ratio of the arc and counterimage in F444W as constraints. Then with those outputs as initial conditions, we performed a second optimization (using a particle swarm optimization algorithm in both instances) over the F444W band image (after subtraction of the foreground elliptical galaxy, see Figure \ref{fig:arc6plex}), including a model for the source light.
Since there is a $\approx 0.5$ mag color gradient between the core and outer region within the lensed arc, we modeled the source light with a point source and two extended components (disk and bulge) using two Sersic profiles. These profiles are constrained to be concentric during the optimization.
The best fit model parameters values obtained after the second optimization are listed in Table \ref{table:summary_param}. The strong lensing modeling was done using the \textsc{lenstronomy}\footnote{https://github.com/sibirrer/lenstronomy} software package (\citealt{birrer2018lenstronomy}, \citealt{Birrer2021}). 

\begin{figure}[ht]
\centerline{
  \includegraphics[ width=1.0\linewidth]{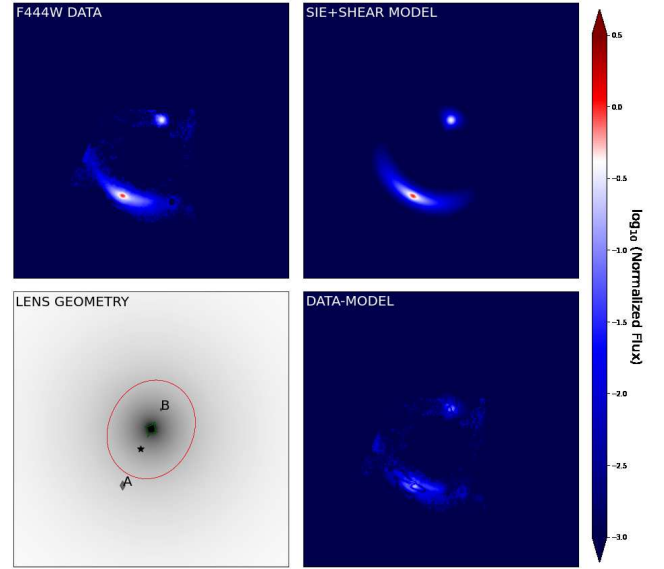} 
}
\caption{(top left) Observed data in band F444W, after subtraction of the elliptical galaxy lens. (top right) Lensing model: a lens with a Singular Isothermal Ellipsoid (SIE) mass distribution and external shear, with an extended source behind it. (bottom left) Lensing geometry reconstruction: the source (black star) is near the cusp of the tangential caustic (green line) and produces two images, of which A is the brightest. A is stretched into an arc due to its proximity to the critical line (red).
(bottom right) Image-Model pixel-by-pixel differences in log scale.
\label{fig:lensingmodel}}
\end{figure}

This model is able to reproduce both the image positions and the flux ratio in the band F444W (see Table \ref{tbl-arcphotometry}). 
The result is shown in Figure \ref{fig:lensingmodel}, along with the lens geometry which requires the source to be near the cusp of the tangential caustic.

Tracing the color of the arc as a function of angle around the lensing nucleus shows a color gradient, with the core 0.44 magnitude redder between F090W and F444W than the surrounding $20^\circ$ away. This might reflect a reddened star-forming core or AGN, where the outer regions of the source are where the outer regions of the source and the central redder component of the source are blended in the counterimage. \changed{This color gradient further motivates our two-component model for the lensed source.} 

The model optimization yields a reconstructed source with a bulge half-light radius of $R_{hl}=0.09"$.
The physical size depends on the angular-diameter distance, and for the arc best fit photo-z source redshift $R_{hl}(z=0.9) \approx 0.75$ kpc. Similarly, the effective velocity dispersion for the isothermal lens model is $\sigma_{SIE} (z=0.9) \approx 250$ km/s which is very close
to the measured velocity dispersion of $\simeq 270\pm7 km/s$ dispersion from the SDSS for VV\,191a, described above, though for a detailed comparison a modest correction for the light profile is required (note that this quantity is not the same as the measured dynamical value \citealt{TreuKoopmans}) because the measured dispersion in the fixed-diameter fiber of SDSS spectroscopy is inherently intensity weighted.
The apparent physical size of the counterimage in the lens plane is $\approx 90$ pc. 

We can take advantage of the symmetry of the lens and the low redshift of the lensing galaxy VV\,191a ($z=0.0513$ corresponding to an angular diameter distance of only 208Mpc) to obtain a precise mass projected within the Einstein radius of 2.01\arcsec. This is model independent, related only to fundamental constants and the distances involved.
The \changed{poorly-known} redshift of the lensed galaxy does not affect this calculation as the lens distance $d_L$
is so small that the distance ratio $d_{LS}/d_S$ is almost unity, for example for the photometric
redshift estimates, the ratio is negligibly different for $z=0.8$ and $z=1.2$ with values of $0.922$ and $0.942$ respectively.
This results in a projected mass within the
Einstein radius of $M(<2.01")\approx1.1\times 10^{11}M_\odot$ and $1\arcsec =1.008kpc$ at the redshift of VV191a.
We can compare this to the projected light from our profile fit for VV\,191a summed within the same $2.01\arcsec$ radius.

To estimate the mass-to-light ratio we convert the HST F606W (which is close to SDSS $r$) luminosity to the $g$ band using the observation that $(g-r)$ colors of low-z luminous elliptical galaxies are always close to 0.80, and use $(g-B)$ transformations from \citealt{Jordi_2006} for elliptical galaxies. \changed{Applying the Galactic foreground extinction estimate from \cite{Schlafly} for the F606W data, we obtain $L_{B}(<2.01") = 1.45 \times10^{10} L_{\odot,B}$, which yields a central mass-to-light ratio of $M/L_B = 7.6 M_\odot/L_{\odot,B}$.}
This value can be compared to the higher-redshift elliptical galaxy lens mass-to-light ratio in the B-band from \cite{TreuKoopmans} of $M/L_B=8 \pm 1 M_\odot/L_\odot$, as well as the results from \cite{Gerhard_2001} for local E/S0 galaxies, which have sample mean values near 7.2 from both dynamical models and individual metallicities plus Salpeter IMF. \changed{Because the lens is comparatively nearby, these values depend only weakly on the distance to the background source. For example, were the source as distant as $z=3$, $M/L_B$ would decrease by 8\%.}

\begin{table}
    \centering
        \begin{tabular}{ |ccccc| }
         
         \hline
         \multicolumn{5}{|c|}{Strong lensing model parameters} \\
         \hline
         \hline
         \multicolumn{3}{|c|}{SIE lens} &\multicolumn{2}{|c|}{SHEAR lens}\\
         \hline
                $\theta_E$ $["]$     & $\phi_{SIE}$ [deg] & $q$  & $\gamma_1$ & $\gamma_2$ \\
         \hline
                $2.01$              & $   -72$ & $0.93$  &  $0.042$          & $-0.075$\\
         \hline 
               \multicolumn{5}{|c|}{Sérsic source}\\
        \hline 
               \multicolumn{2}{|c|}{Disk} &\multicolumn{2}{|c|}{Bulge} &\\
         \hline
                $R_{S}$ $["] $     & $n_{S}$ $(^*)$ &$R_{S}$ $["] $     & $n_{S}$ $(^*)$ & center$_{src}$\\
         \hline
                0.21 & 1 & 0.09 & 4 & ($-0.436, -0.854$)\\
         \hline 
        \end{tabular}
    \caption{Best fit lens and source component parameters. The SIE component is characterized by the Einstein radius $\theta_E$, the inclination angle $\phi_{SIE}$, and the axial ratio $q$. The shear is defined by the two components $(\gamma_1, \gamma_2)$. The extended source is modeled with two Sérsic profiles, defined by the half-light radius $R_{S\acute{e} rsic}$, the power-law index $n_{S\acute{e} rsic}$ and the center in the source plane. The parameters marked with $(^*)$ are kept fixed during the optimization. The center of the lens components is kept fixed at $(0, 0)$.}
    \label{table:summary_param}
\end{table}

\noindent

\begin{deluxetable*}{lccccc}
\tablecaption{Photometry of the VV191a Possible Lensing Features} 
\label{tbl-arcphotometry}
\setlength{\tabcolsep}{7pt}
\tablehead{
\colhead{Measurement} & \colhead{WFC3} & \colhead{NIRCAM} & \colhead{NIRCAM} & \colhead{NIRCAM} & \colhead{NIRCAM} \\[-8pt]
 & \colhead{F606W} & \colhead{F090W} & \colhead{F150W} &
 \colhead{F356W} & \colhead{F444W}
}
\startdata
{\bf Arc integrated:} \\
AB mag & $23.16\pm 0.3$ & $20.71\pm 0.05$ &	$19.73\pm 0.05$ & $19.03\pm 0.05$ & $19.39\pm 0.05$  \\ 
{\bf Arc core:}\\
AB mag & $23.65\pm 0.15$ & 	$21.16\pm 0.05$ & 	$19.95\pm 0.05$ &	$19.41\pm 0.05$ & 	$19.77\pm 0.05$ \\ 
{\bf Counterimage CI:}\\
Arc/counterimage ratio & --- & 8.09 &	9.63 & 6.92 & 6.49 \\ 
AB mag &  -- & $22.98\pm 0.05$ & $21.89\pm 0.05$ & $21.13\pm 0.05$ &	$21.42\pm 0.05$  \\ 
{\bf Weak lens candidate A:}\\
AB mag & --- & $25.80\pm 0.20$ & $22.94\pm 0.05$ & $21.44\pm 0.05$ &	$21.26\pm 0.05$ \\ 
\bf{Diffuse component:}\\
AB mag &  --- &  $23.51\pm 0.25$ & $22.52\pm 0.18$ & $22.11\pm 010$ &	$22.63\pm 0.10$ \\ 
[2pt]
\enddata	 
\end{deluxetable*}

The image configuration is quite similar to that seen in the strongly magnified systems IRAS F10214+4724 \citep{IRAS10214} and 3C 220.3 \citep{Haas2014}. In exploring mass and lensing models, we note the SDSS DR15 velocity dispersion $272 \pm 7$ km s$^{-1}$ with the BOSS fiber aperture of diameter 2\arcsec.

An additional faint red galaxy appears \changed{at right} in Figure \ref{fig:arc6plex}, 5.25\arcsec from the foreground nucleus. This object is elongated tangential to the foreground galaxy (best seen in F444W), in a region where the arc model predicts that weak lensing should stretch the image by about a factor 1.6 \changed{without generating a counterimage, so this is a second plausibly lensed background system. It is also very red; the F356W and F444W magnitudes are more reliable because the F090W and F150W images are confused with a blue star image $0.18$\arcsec away (which is too faint to be a major contaminant in the longer bands). \changed{Uncertainty due to deblending star and galaxy fluxes} is indicated by colons for these passbands in Table \ref{tbl-arcphotometry}. EAZY gives a primary photometric redshift $z=2.06\pm 0.10$ (55\% likelihood) with a secondary (34\%) peak at $z=1.7 \pm 0.1$.}

\changed{A similar EAZY fit gives photometric redshift estimate
$z=0.73 \pm 0.18$ for the diffuse component outside the western end of the lensed arc, allowing but not requiring its source to be
physically associated with that of the arc.}


\bigskip


Galaxy-galaxy gravitational lensing was a later discovery than lensing by groups and clusters, since the Einstein radii are smaller and the light of the lensed source may be blended with that from the lens galaxy. However, the more detailed work needed to find such lenses has paid off in tracing the mass distributions of (especially early-type) galaxies over a substantial redshift range, with detection techniques ranging from high-redshift AGN spectra in low-redshift galaxies \citep{Huchra85}, through multiple redshift systems in SDSS spectra (the SLACS survey, \citealt{SLACS04,SLACS05,SLACS06}) and searches for lens structures in HST images (a recent summary of the many advances in finding such lens systems is given by \citealt{DESBAS}). The difficulty of recognizing the lensed components in VV191 from even HST data in the optical, together with their prominence in the NIRCam images, suggests that large numbers of galaxy-galaxy lenses will appear (even serendipitously) in JWST galaxy studies. Such samples will improve our understanding of the mass distribution in the lens galaxies over cosmic time, in favorable cases also telling us about the masses of supermassive black holes. In VV\,191a, the lack of a detectable third image (not seen either in the data we show here or the HST near-UV data) is consistent with the SMBH masses typically expected at this luminosity, as is the mass model.
\bigskip

\section{Summary and Conclusions}

Combining HST and JWST imaging, we have used the
backlighting of the face-on spiral galaxy VV\,191b by a luminous elliptical to trace the structure and reddening law of dust in its outer disk. The unusually favorable geometry of this system allows us to map the dust attenuation with spatial resolution $\approx 60$ pc, spanning radii 12-21 kpc (much of this range, beyond the stellar spiral arms), and with high-quality mapping from 0.6-1.5$\mu$m. 

The distinct dust lanes end sharply at radius 21 kpc (1.7 Petrosian radii), and exhibit not only the usual pattern following the pitch angle of the bright spiral arms but additional lanes cutting at a variety of angles to this.

We fit curves for various values of total-to-selective extinction $R$ to the PSF-matched attenuation values measured pixel-by-pixel in multiple filters from 0.6-1.5~\micron, to estimate the slope of the effective reddening law. Most regions can be described by \changed{$R\approx 2.0$ using a CCM reddening law, or with power-law index $\beta \approx 2.0$}, with longer wavelengths showing
smaller $R$ \changed{and more positive $\beta$ (less-gray} extinction). \changed{Limits to the attenuation at longer wavelengths 3.5-4.4$\mu$m are 0.02-0.035 $A_V$.} Although even sub-percent systematics in modeling the galaxy brightness profiles are important for small attenuation values, we see some evidence that small-attenuation regions have larger $R$ than regions with more attenuation, which could be a sign that clumpy structure affects strong-attenuation areas more strongly. \changed{The most striking feature of these measurements is how strongly reddening the behavior is, even in the likely presence of mixing effects combining region of different attenuation in each resolution element, which we in fact detect by measuring our transmission maps after various degrees of averaging. Broadly, this might suggest a population strongly dominated by small grains, such as has been suggested in the outer disk of the Milky Way by \cite{Zasowski} based on stellar colors in the near-infrared. It may be important in shaping the grain population the the dust features we sample in VV\,191b are as far as 7.5 kpc from regions of ongoing star formation as traced in the UV or by hot dust.}

\changed{This investigation of dust far from star-forming regions may be relevant to the
extinction and reddening in cosmologically-interesting SN Ia, which may explode anywhere in
a galaxy (in contrast to the concentration of core-collapse supernova near their
formation regions). Supernovae probe extinction much more precisely than our
spatially-averaged measurements, approaching the point-source probes provided by ordinary stars.
\cite{ThorpMandel} use a hierarchical Bayesian approach to address the possibility of
unphysical ranges of $R$ retrieved in supernova samples due to observational error,
finding $R=2.2-2.6$ for type Ia supernova in the Carnegie Supernova project. They
contrast this with $R=1.9$ from fits to individual SN from \cite{Johansson}. All
the values are significantly lower than the frequently-used Milky Way (local)
mean $R\approx 3.5$, in the same sense as our results for the outer disk of VV\,191b.}

The high quality of the NIRCam data from 0.9 to 1.5~\micron\  means that the NIR is now the most sensitive regime to trace backlit dust in galaxies, even with its greater attenuation at shorter wavelengths. These observations detect dust lanes farther out than the spiral pattern is seen in starlight even with the deep IR images. F150W is surprisingly valuable for its leverage in fitting the slope of  the effective reddening law (F070W+F150W would be a fast survey pair with NIRCam, with longer wavelengths observed at no extra cost in observing time). The Galaxy Zoo overlapping-galaxy catalog alone \citep{GZcatalog} includes a number of deeper overlap pairs within the distance $z=0.05$ of this one, which would be at least as well resolved and allow construction of composite attenuation profiles which could be valuable in constructing priors for supernova photometry into the near-IR, and improving our understanding of the energy budget in disk galaxies. Complementing such efforts, these optical/NIR attenuation maps could provide a comparison set for data tracing the PAH emission bands farther into the infrared, and far-IR or submillimeter tracers of the overall dust emission, to refine our understanding of how the components of the grain population change in various environments.

The JWST images further revealed a galaxy-galaxy gravitational lensing instance through the elliptical galaxy VV\,191a, with a background system at \changed{$z \approx 0.9$} appearing as a $90^\circ$ arc and counterimage. An additional distant background galaxy appears weakly lensed with modeled axial ratio 1.6. All these can be reproduced simultaneously with a lens modeling including a singular isothermal sphere and shear (likely associated with the adjacent spiral galaxy). The small redshift of the lens allows a determination of the mass/light ratio within its Einstein radius almost independent of the source redshift, $M/L_B=7.6$ in solar units.

VV\,191 vividly demonstrates a facet of backlit-galaxy dust analysis which becomes important only at significant redshifts: large redshift differences increase the role of gravitational lensing in distorting the background sources, so using galaxy symmetry in modeling is more effective for small redshift offsets between the two galaxies.

\clearpage

\begin{acknowledgements}

We dedicate this paper to Prof. Hendrik van de Hulst, who during his long and very productive career at Sterrewacht Leiden had a mission to understand cosmic dust.
This project is based on observations made with the NASA/ESA {\emph Hubble Space Telescope} and the NASA/ESA/CSA \emph{James Webb Space Telescope}, obtained from the Mikulski Archive for Space Telescopes, which is a collaboration between the Space Telescope Science Institute (STScI/NASA), the Space Telescope European Coordinating Facility
(ST-ECF/ESA), and the Canadian Astronomy Data Centre (CADC/NRC/CSA). \changed{The IUE spectra were retrieved from the
INES service of LAEFF in Spain.}

RAJ, RAW, and SHC acknowledge support from NASA \JWST\ Interdisciplinary
Scientist grants NAG5-12460, NNX14AN10G and 80NSSC18K0200 from GSFC.
CNAW acknowledges funding from the JWST/NIRCam contract NASS-0215 to the
University of Arizona.  We acknowledge support from the ERC Advanced Investigator Grant EPOCHS (788113), as well as a studentship from STFC. AZ acknowledges support by Grant No. 2020750 from the United States-Israel Binational Science Foundation (BSF) and Grant No. 2109066 from the United States National Science Foundation (NSF), and by the Ministry of Science \& Technology, Israel.

This research was supported by the Australian Research Council Centre of Excellence
for All Sky Astrophysics in 3 Dimensions (ASTRO 3D), through project number CE170100013.

We thank our Program Coordinator, Tony Roman, for his expert help scheduling this program.
We thank the Galaxy Zoo volunteers for finding so many of these backlit-galaxy systems (particularly Galaxy Zoo users \changed{kodemunkey and} 
Goniners for first pointing out this pair).

\end{acknowledgements}

%

\software{
	\code{MultiDrizzle/DrizzlePac} \citep{MultiDrizzle,AstroDrizzle},
	\code{Python(AstroConda/Astropy)} \citep{Astropy2013,Astropy2018}:
              \url{http://www.astropy.org},
        \code{Matplotlib} \citep{Matplotlib2007},
        \code{IRAF}, \code{IDL},
        \code{lenstronomy} \citep{birrer2018lenstronomy,Birrer2021},
        \code{EAZY} \citep{EAZY}.
}

\facilities{James Webb Space Telescope (Near InfraRed Camera); Hubble Space Telescope (Wide Field Camera 3)
Mikulski Archive for Space Telescopes:
\url{https://archive.stsci.edu}}

The data presented in this paper were obtained from the Mikulski Archive for Space 
Telescopes (MAST) at the Space Telescope Science Institute. The specific observations analyzed can be 
accessed via \dataset[10.17909/78wg-t688]{https://doi.org/10.17909/78wg-t688}.

\end{document}